\documentclass[a4paper,11pt]{article}
\pdfoutput=1 
\usepackage{jheppub} 
\usepackage[T1]{fontenc} 
\usepackage{comment}
\usepackage{amsmath}
\usepackage{caption}
\usepackage{subcaption}
\usepackage{float}
\usepackage{enumitem}
\usepackage[normalem]{ulem}
\usepackage[dvipsnames]{xcolor}
\usepackage{pifont}
\usepackage{longtable}
\usepackage{cancel}
\usepackage[stable,bottom]{footmisc}

\usepackage{array}

\usepackage{cleveref}
\crefname{figure}{Fig.}{Figs.} 
\crefname{section}{Sec.}{Secs.}

\usepackage{tikz}
\usetikzlibrary{calc}
\usepackage{circuitikz}

\usepackage{textcomp}
\usepackage{bm}

\usepackage{makecell}

\newcommand{\omegat}{\bar{\omega}}

\newcommand{\x}{\chi}

\newcommand{\mn}{{\mu \nu}}
\newcommand{\ab}{{\alpha\beta}}

\newcommand{\eq}{0}
\newcommand{\const}{{\rm const}}

\newcommand{\pars}[1]{\left( #1 \right)}

\newcommand{\spars}[1]{\left[ #1 \right]}

\newcommand{\thev}[1]{\left\langle#1\right\rangle}
\newcommand{\ind}[1]{{\scalebox{.6}{$ #1 $}}}

\newcommand{\rr}{\rho}
\newcommand{\V}{\mathcal{V}}

\title{The probe limit in MHD and its implications for magnetic transport
}

\author[a]{Giorgio Frangi,}
\author[b]{Matej Bajec,}
\author[b]{Guri K.~Buza,}
\author[b]{Alexander Soloviev,}
\author[a,b]{and \\Sa\v so Grozdanov}

\affiliation[a]{Higgs Centre for Theoretical Physics, School of Physics and Astronomy, University of Edinburgh,
Peter Guthrie Tait Road, King’s Buildings, Edinburgh EH9 3FD, Scotland}
\affiliation[b]{Faculty of Mathematics and Physics, University of Ljubljana, Jadranska ulica 19, SI-1000, Ljubljana, Slovenia}

\abstract{Many phenomenological and effective field-theoretical (EFT) applications of magnetohydrodynamics (MHD) in the presence of a background magnetic field employ a simplifying assumption whereby the electromagnetic and the energy-momentum fluctuations decouple. In studies of magnetic transport, for example in magnetic diffusion, the conservation of energy and momentum is then neglected. In this paper, we investigate the details and the consistency of this so-called {\em probe limit} in different parametric regimes of MHD plasmas. In the first part of the paper, our discussion explores the hydrodynamic (higher-form) theory of MHD. In the second part, we then explicitly test the probe limit by using a microscopic holographic (AdS/CFT) model of a strongly coupled plasma. In the process, we develop the holographic Schwinger-Keldysh EFT prescription for describing the bulk 2-form fields and their dual 1-form symmetries. Moreover, we find evidence of a phase transition at low temperatures and show that magnetic Hall transport can emerge as a consequence of background charge density that breaks the charge conjugation symmetry of the state. Finally, we discuss the implications for magnetic transport, with a particular view towards the dynamics of dense nuclear matter in neutron stars.}

\begin{document}
\maketitle

\newpage

\section{Introduction}

From a modern perspective, hydrodynamics is the universal effective field theory (EFT) describing the late-time, long-distance evolution of conserved densities of a physical system (see e.g.~\cite{Kovtun:2012rj,Dubovsky:2011sj,Grozdanov:2013dba,Crossley:2015evo,Haehl:2018lcu}). Its well-posedness as an EFT hinges on the existence of a clear timescale separation between microscopic interactions and macroscopic transport phenomena, encapsulated in an infrared (IR) derivative expansion. This approach, based on the underlying symmetries of the theory and the state, has led to significant progress in generalizing and understanding transport in many IR phases of matter (among many such recent papers, see e.g.~\cite{Grozdanov:2016tdf,Gralla:2018kif,Glorioso:2018kcp,Delacretaz:2016ivq,Grozdanov:2018ewh,Armas:2019sbe,Delacretaz:2019brr,Baggioli:2020haa,Amoretti:2021fch,Landry:2021kko,Baggioli:2021ntj,Jain:2021ibh,Baggioli:2022pyb,Armas:2022vpf,Iqbal:2020lrt,Armas:2023tyx,Davighi:2023mzg} and references therein).

A physical system may have several conserved currents that correspond to its underlying continuous global symmetries. In this case, their dynamics is generically coupled. Assume the (relativistic) energy-momentum tensor $T^\mn$ to be one of them. In terms of the effective hydrodynamic variables (the temperature $T$, fluid velocity $u_\mu$ and any chemical potentials associated to internal symmetries, collectively denoted as $\mu_I$), this amounts to a presumption that by perturbing the system slightly out of equilibrium, any nontrivial solution to the conservation equations should involve all of them. It is in special cases only -- for instance when the equilibration timescale of $T^\mn$ is much longer than the one of internal currents, in turn much longer than any microscopic one -- that some of the perturbations can be effectively switched off. In those cases, one can find a solution to the conservation equations such that:
\begin{equation}
    \label{eq:pbdef}
    \delta T,\delta u^\mu = 0, \quad \delta \mu_I \neq 0.
\end{equation}
This regime is referred to in the literature as the \textit{probe limit} of hydrodynamics. Its utility and limitation can be illustrated by considering the case of the IR dynamics of a conserved ordinary $U(1)$ current $j^{\mu}$, coupled with $T^\mn$. If the equilibrium thermal state that is being perturbed has a `baryonic' number density $n_0=0$, there is a nontrivial solution with $\delta T, \,\delta u=0$ corresponding to simple charge diffusion. No such fully decoupled and consistent solution exists if $n_\eq\neq0$, reflecting the fact that the conserved densities evolve concurrently. 

Despite it being formally inconsistent, the probe limit may still capture the relevant physics of interest in certain parametric regimes of the theory. In fact, when studied from the point of view of AdS/CFT, the full theory of charged hydrodynamics with a single $U(1)$ current is typically taken to be dual to the Einstein-Maxwell theory in the charged Reissner-Nordstr\"{o}m black brane background. It should not be overlooked that from the holographic point of view, this too is an inconsistent probe limit of the top-down string theoretic construction. As was recently pointed out, the presumably consistent stringy solution known as the STU black hole \cite{Behrndt:1998jd}, which gives rise to R-charged hydrodynamics \cite{Son:2006em} involving three chemical potentials and three $U(1)$ conserved currents, cannot be smoothly deformed to give the Einstein-Maxwell theory with the Reissner-Nordstr\"{o}m black hole. Before the parameters can be tuned to allow for a single chemical potential, the full theory hits an instability \cite{Gladden:2024ssb}. In the broader context, this is simply a cautionary tale that any probe limit should be understood in an effective context that suppresses some part of the physical spectrum, and that such a limit may or (more usually) may not be a consistent truncation of the original theory. 

In this work, we scrutinize the probe limit of relativistic magnetohydrodynamics (MHD), using the higher-form formalism developed in \cite{Grozdanov:2016tdf} (see also \cite{Hernandez:2017mch,Armas:2018atq} and subsequent works). This formulation of MHD is naturally apt to describe electromagnetic and thermal transport in presence of strong magnetic fields, with concrete applications to dense nuclear matter in neutron stars \cite{Vardhan:2022wxz}. In this context, the probe limit \eqref{eq:pbdef} has played an important part, including in the seminal phenomenological work by Goldreich and Reisenegger \cite{goldreich1992magnetic} -- upon which many variations of MHD equations have been built in the astrophysical context. More recently, two works \cite{Vardhan:2022wxz,Vardhan:2024qdi} generalized its results by adopting modern Schwinger-Keldysh EFT techniques along with the language of higher-form symmetries to derive and classify different types dissipative effective action for MHD, precisely in the probe limit of the 2-form current discussed here. In this work, we perform a detailed comparison of the modes, correlation functions and various predictions that come from the full theory of MHD and its probe limit, so to develop a better understanding of the latter.

The structure of the paper is as follows. In \Cref{sec:mhd} we derive the full set of correlators of MHD from its conservation equations in presence of a background magnetic field, and provide explicit Kubo formulae for its transport coefficients. We then show that any nontrivial probe limit solution \eqref{eq:pbdef} is approximate and that it strictly violates the conservation of $T^\mn$. To substantiate these claims, we calculate the transport coefficients in a microscopic holographic theory of plasma \cite{Grozdanov:2017kyl,Hofman:2017vwr}, both in the probe limit -- which in gravity corresponds to freezing metric perturbations -- and in the full theory with a metric and a bulk massless 2-form gauge field that sources the 2-form current on the boundary. In \Cref{sec:grav_probe}, we construct a probe limit effective action using the holographic Schwinger-Keldysh formalism, and find exact expressions for the transport coefficients in terms of the background metric. In the subsequent \Cref{sec:beyond}, we compare these predictions against the results obtained by calculating the correlators in the corresponding fully backreacting geometry, and find perfect agreement with our earlier claims. As an offshoot, in \Cref{sec:beyond}, we widen the scope of our investigation to include a finite background charge density $n_\eq$ in the analysis, which induces a background electric field. The main purpose is to show that, as recently discussed in \cite{Vardhan:2022wxz,Vardhan:2024qdi}, the breaking of the charge conjugation symmetry should lead to the appearance of a third magnetic transport coefficient -- the Hall resistivity $r_H$ (see also \cite{Hernandez:2017mch}). This fills a gap in the AdS/CMT literature regarding the appearance of a Hall resistivity in a 4$d$ boundary theory, a phenomenon which had previously been studied exclusively in lower dimensions \cite{Hartnoll:2007ai}. In \Cref{sec:neutron}, we use the transport coefficients calculated from holography as an input to discuss phenomenological differences between the various regimes of magnetic transport considered in this and other works. We then conclude by outlining a few possible extensions of this work in \Cref{sec:disc}. Three appendices have been added to provide further technical details for the interested reader. 

\paragraph{Notation and conventions}--- Throughout this work, we reserve lowercase Greek letters ($\mu,\nu,...$) for spacetime indices, lowercase Latin ($m,n,...$) for purely spatial indices, and uppercase Latin ($M,N,...$) for spacetime indices in the holographic bulk (\Cref{sec:grav_probe,sec:beyond}). Any deviations from this convention are appropriately marked in the text. Time indices are always denoted by $t$ (or the Eddington-Finkelstein $v$ only in the holographic bulk of \Cref{sec:grav_probe}). The number 0 as a subscript is used to denote equilibrium thermodynamic quantities, which we drop from \Cref{sec:beyond} onward. Our convention for the Fourier transform is 
$$f(\omega,\mathbf k) = \int_{-\infty}^\infty dt \int d^3 \mathbf x\, e^{i\omega t-i\mathbf k \cdot \mathbf x}f(t,\mathbf x).$$
(Anti)-symmetrization is defined as:
$$X^{[ab]} = \frac{1}{2}(X^{ab}-X^{ba}), \quad X^{(ab)} = \frac{1}{2}(X^{ab}+X^{ba}).$$
Throughout this text we use $\rho$ to denote the density of magnetic field lines (which corresponds to the magnetic field intensity $B$ adopted in many works), and $n$ for conserved charge density.

Lastly, an important distinction in the following is the one between \textit{magnetic viscosities} (denoted as $r_\parallel,~r_\perp$ and $r_H$) and \textit{electrical resistivities} (the components of the thermomagnetic matrix $r^{ij,kl}$). In probe limit calculations, the two are effectively indistinguishable (cf.~\eqref{eq:visc=res}), but in the full theory, they are not (cf.~\eqref{eq:tmtrcfs}). Though the terms are used somewhat interchangeably, we will make distinctions whenever appropriate.

\section{Magnetohydrodynamics}
\label{sec:mhd}

In this work, we adopt the viewpoint presented in \cite{Grozdanov:2016tdf} and formulate magnetohydrodynamics (MHD) as a hydrodynamic EFT of a conserved 2-form current $J^\mn$ coupled with the energy-momentum tensor $T^\mn$. The 2-form equation enshrines magnetic field flux conservation and is in effect a consequence of the Bianchi identity, $dF=0$, valid in any theory with dynamical electromagnetism without magnetic monopoles. Its corresponding symmetry is a 1-form $U(1)$ symmetry \cite{Gaiotto:2014kfa}, and we denote its external source, a 2-form gauge field, as $b_\mn$. The conservation equations for this theory are:
\begin{subequations}
\label{eq:conservation}
\begin{align}
    \nabla_\mu J^\mn &= 0,\label{eq:Jmn_Conservation}\\
    \nabla_\mu T^\mn &= H^\nu_{\phantom{\nu}\ab} J^\ab \label{eq:Tmn_Conservation},
\end{align}
\end{subequations}
where the tensor strength $H=db$ is expressed in coordinates as
\begin{align}
    H_{\mu\alpha\beta}= \partial_\mu b_\ab+\partial_\beta b_{\mu\alpha}+\partial_\alpha b_{\beta\mu}.
\end{align}
Besides the temperature $T$ and fluid velocity $u^\mu$, the other fundamental hydrodynamic variables are the chemical potential $\mu$ (conjugated with the magnetic field line number density $\rr$ directly related to the magnetic field $B$) and the spacelike vector field $h^\mu$ which points in the direction of the magnetic field. The normalization and orthogonality conditions, $u^\mu u_\mu = -1$, $h^\mu h_\mu = 1$ and $u^\mu h_\mu = 0$, reduce the number of independent degrees of freedom to seven, as many as there are dynamical equations of motion. 

\bgroup
\def\arraystretch{1.3}
\begin{table*}[t!]
\footnotesize
\centering
\begin{tabular}{|cccccccc|}
\hline
 & $J^{ti}$  & $J^{ij}$  & $u^{t}$ & $u^{i}$ & $h^{t}$ & $h^{i}$ & $\rho,\mu,\varepsilon,p$ \\\hline
$C$ & -- & -- & + & + & -- & -- & + \\
$P$ & + & -- & + & -- & -- & + & + \\
$T$ & -- & + & + & -- & + & -- & + \\
\hline
\end{tabular}
\caption{Transformation rules of $J^\mn$ and hydrodynamic degrees of freedom under the fundamental discrete symmetries: $C,~P$ and $T$.}
\label{tab:disc_symm}
\end{table*}
\egroup

Guided by the underlying symmetries of our theory, we next construct the constitutive relations for $T^\mn$ and $J^\mn$. In doing so, we must assign to each variable specific transformation properties under discrete symmetries, which we report in \Cref{tab:disc_symm}. For additional details, see \cite{Grozdanov:2016tdf,Vardhan:2024qdi}. Then, the constitutive relations can be systematically organized within a gradient expansion. To first order in gradients, we have \cite{Grozdanov:2016tdf}
\begin{subequations}
\label{eq:ConstRel}
\begin{align}
    J^\mn &= 2\rr u^{[\mu}h^{\nu]} + 2 m^{[\mu}h^{\nu]}+s^\mn,\label{eq:Jmn_ConstRel}\\
    T^\mn &= (\varepsilon+p)u^\mu u^\nu +  pg^\mn - \left(\mu\rr -\delta \tau \right) h^\mu h^\nu +\delta f \Delta^\mn+ 2\ell^{(\mu}h^
    {\nu)}+t^\mn,\label{eq:Tmn_ConstRel}
\end{align}
\end{subequations}
where $\Delta^\mn = g^\mn + u^\mu u^\nu - h^\mu h^\nu$ is the projector perpendicular to both $u$ and $h$, and
\begin{subequations}
\label{eq:tfapp}
\begin{align}
& \ell^{\mu} = -2\eta_{\parallel}\Delta^{\mu\sigma}h^{\nu} \nabla_{(\sigma}u_{\nu)} ,\label{visc4} \\
& t^{\mu\nu} = -2\eta_{\perp}\left(\Delta^{\mu\rr}\Delta^{\nu\sigma}- \frac{1}{2} \Delta^{\mu\nu}\Delta^{\rr\sigma}\right)\nabla_{(\rr}u_{\sigma)} ,\\
& m^{\mu}  = -2 r_{\perp}\Delta^{\mu\beta}h^{\nu}\left( T \nabla_{[\beta}\left(\frac{h_{\nu]} \mu}{T}\right) + u_{\sigma} H^{\sigma}_{\phantom{\sigma}\beta\nu}\right) ,\label{mdef}\\
& s^{\mu\nu} = -2 r_{\parallel}\Delta^{\mu\rr} \Delta^{\nu\sigma} \left(\mu \nabla_{[\rr} h_{\sigma]} +H^{\lambda}_{\phantom{\sigma}\rr\sigma}u_{\lambda}\right)\label{sdef},\\
& \delta f = -\zeta_{\perp} \Delta^{\mu\nu} \nabla_{\mu}u_{\nu} - \zeta_{\times} h^{\mu}h^{\nu} \nabla_{\mu}u_{\nu} \label{bulk1}, \\
& \delta \tau  = -\zeta_{\times} \Delta^{\mu\nu} \nabla_{\mu} u_{\nu} - \zeta_{\parallel} h^{\mu} h^{\nu} \nabla_{\mu} u_{\nu} \label{viscfin}.
\end{align}
\end{subequations}
The dispersion relations following from the linearized theory have been obtained and studied in \cite{Grozdanov:2016tdf}. For convenience we report them here. Parameterizing spatial momentum as $k^i = \kappa\left(\sin\theta,0,\cos\theta\right),$ we find Alfvén waves in its transverse plane:
\begin{subequations}
\begin{equation}
    \omega = \pm v_A  \kappa - \frac{i}{2}\Gamma_A \kappa^2,\label{eq:Dispersion_Alfven} 
\end{equation}
and magnetosonic waves in its parallel direction:
\begin{equation}
    \omega  = \pm v_M \kappa - i \tau \kappa^2.\label{eq:Dispersion_Magnetosonic}
\end{equation}
\end{subequations}
The speeds of sound are given by:
\begin{subequations}
\begin{align}
    &v_A^2 = \frac{\mu_\eq\rho_0}{\varepsilon_0+p_0}\cos^2\theta \equiv \V_A^2 \cos^2\theta,\\[4pt]
    & v_M^2 = \frac{1}{2} \bigg\{ 
\left( \mathcal{V}_A^2 + \mathcal{V}_0^2 \right) \cos^2 \theta 
+ \mathcal{V}_S^2 \sin^2 \theta 
\nonumber\\
&\qquad\qquad \pm 
\sqrt{ \left[ (\mathcal{V}_A^2 - \mathcal{V}_0^2) \cos^2 \theta + \mathcal{V}_S^2 \sin^2 \theta \right]^2 
+ 4 \mathcal{V}^4 \cos^2 \theta \sin^2 \theta } 
\bigg\},
\end{align}
\end{subequations}
and we defined
\begin{align}
    \V^2_0 &= \frac{s_0\chi}{T_\eq(c\chi-\lambda^2)},\quad
    \V_S^2 =\frac{s_0^2\chi+\rho^2c-2\rho_0\lambda}{(c\chi-\lambda^2)(\varepsilon_0+p_0)}, \quad
    \V^4 =\frac{s(\rho_0\lambda-s_0\chi)^2}{T_\eq(c\chi-\lambda^2)^2(\varepsilon_0+p_0)},
\end{align}
and
\begin{align}
    \chi = \left(\frac{\partial \rho}{\partial \mu}\right)_T,\quad \lambda=\left(\frac{\partial \rho}{\partial T}\right)_\mu,\quad c=\left(\frac{\partial s}{\partial T}\right)_\mu.
\end{align}
The attenuation of Alfvén waves in \eqref{eq:Dispersion_Alfven} is given by
\begin{align}
    \frac{1}{2}\Gamma_A \kappa^2=k_i\left(\gamma^{ij}_\eta+ \gamma^{ij}_r\right)k_j,
\end{align}
where 
\begin{align}
    \gamma_\eta &= (\varepsilon_0+p_0)^{-1}\,\text{diag}\left(\eta_\perp, \eta_\parallel\right), \nonumber \\
   \gamma_r  &= (\rho_0/\mu_\eq)^{-1}\,\text{diag}\left( r_\parallel, r_\perp \right).\label{eq:gammas}
\end{align} As already suggested in \cite{Grozdanov:2016tdf}, the expression for $\tau$ in \eqref{eq:Dispersion_Magnetosonic} is complicated, so we will not report it here and refer to that work for a few limiting cases of interest.

\paragraph{A note on the number of magnetic viscosities}--- It is apparent from \eqref{eq:tfapp} that only two magnetic viscosities ($r_\parallel$ and $r_\perp$) appear in our theory. This is because of the choice of the $C$, $P$ and $T$ symmetries under which the constitutive relations are invariant. A way to introduce the third, Hall magnetic viscosity $r_H$ is to break the charge conjugation $C$ symmetry \cite{Vardhan:2022wxz,Vardhan:2024qdi} (see also 
\cite{Hernandez:2017mch}). We will further discuss $r_H$ in \Cref{sec:beyond}.

\subsection{Full linearized theory in variational approach}\label{sec:full_MHD_Var}

Next, we outline the derivation of correlators arising from the full theory of MHD, mirroring the discussion in~\cite{Kovtun:2012rj} and working in the variational approach. 

From here onward we consider hydrodynamic fluctuations of a thermal state with a uniform background (dynamical) magnetic field. Without loss of generality, we work in the rest frame of the background fluid and in it align the magnetic field along $z$. All in all, this corresponds to taking
\begin{equation}
	\label{eq:hveq}
	T_\eq = \const., \quad \mu_\eq = \const., \quad u^\mu_\eq = \delta^{\mu}_t, \quad h^\mu_\eq = \delta^\mu_z.
\end{equation}
We choose the background metric to be flat Minkowski spacetime, $\eta_\mn$, and the background gauge field to be of the form $b_{\eq}^\mn = \mu_\eq u_{\eq}^{[\mu} h_{\eq}^{\nu]}$. In equilibrium, the expressions for $T^\mn$ and $J^\mn$ (see \eqref{eq:ConstRel}) reduce to
\begin{subequations}
\label{eq:ceqneq}
\begin{align}
    J^\mn_\eq&=2\rr_0u^{[\mu}_\eq h^{\nu]}_\eq, \\
    T^\mn_\eq &= (\varepsilon_0+p_0)u^\mu_\eq u^\nu_\eq+p_0\eta^\mn -\mu_\eq\rr_0 h^\mu_\eq h^\nu_\eq,
\end{align}
\end{subequations}
which identifies $\varepsilon_0$, $p_0$ and $\rr_0$ with equilibrium energy density, pressure, and density of magnetic field lines along the $z$ direction. 

Next, we perturb all of the macroscopic variables around their equilibrium value \eqref{eq:hveq}:
\begin{align}
	& T(t,\mathbf{x}) = T_\eq + \delta T(t,\mathbf{x}), \quad  u^\mu(t,\mathbf{x}) = u^\mu_\eq + \delta u^\mu(t,\mathbf{x}),  \nonumber \\  & \mu(t,\mathbf{x}) = \mu_\eq + \delta \mu(t,\mathbf{x}), \quad \, h^\mu(t,\mathbf{x}) = h^\mu_\eq + \delta h^\mu(t,\mathbf{x}) \label{eq:hvpt},
\end{align}
and obtain:
\begin{subequations}
\begin{align}
\varepsilon(t,\textbf{x})&=\varepsilon_0+\delta \varepsilon(t,\textbf{x})
=\varepsilon_0 +\left(\frac{\delta \varepsilon}{\delta T}\right)_\mu \delta T+ \left(\frac{\delta \varepsilon}{\delta \mu}\right)_T \delta \mu,\\
 p(t,\textbf{x})&=p_0+\delta p(t,\textbf{x})=p_0 +\left( \frac{\delta p}{\delta T}\right)_\mu \delta T+\left( \frac{\delta p}{\delta \mu} \right)_T \delta \mu,\\
    \rr(t,\textbf{x})&=\rr_0+\delta \rr(t,\textbf{x})=\rr_0 +\left(\frac{\delta \rr}{\delta T}\right)_\mu \delta T+ \left( \frac{\delta \rr}{\delta \mu}\right)_T \delta \mu.
\end{align}
\end{subequations}
Using the Gibbs-Duhem relation $dp = sdT+\rho d\mu$, we note
\begin{align}
    \left( \frac{\delta p}{\delta T}\right)_\mu=s_0\quad\text{and}\quad \left(\frac{\delta p}{\delta \mu}\right)_T=\rho_0,
\end{align}
where $s_0$ is the equilibrium entropy density, given by the first law of thermodynamics $\varepsilon+p=sT+\mu\rr$. 

In the variational approach, the hydrodynamic perturbations \eqref{eq:hvpt} are induced by the perturbations of the metric tensor and the 2-form gauge field
\begin{equation}
    g_\mn=\eta_\mn+\delta g_\mn, \quad b_\mn = b^\eq_\mn +\delta b_\mn.
\end{equation}
To obtain the retarded correlators, one must solve the (linearized) hydrodynamic equations of motion \eqref{eq:conservation} in Fourier space for the fluctuating hydrodynamic variables $\delta T$, $\delta \mu$, $\delta u^i$ and $\delta h^i$. A naive counting of the equations of motion might lead us to think that there is one too many. However, not all four equations in \eqref{eq:Jmn_Conservation} are independent, and in this work, we remove its $\nu=t$ component, which acts as the magnetic Gauss's law constraint on the initial conditions.\footnote{We note that this constraint is satisfied by our solution. Alternatively, one can remove, in the linear regime, any of the three equations which lie in the space spanned by our chosen $k^\mu$. For the case at hand, this is any of the $t$, $x$ or $z$ equations.} 

The solution to the linearized equations of motion yields the expressions for the hydrodynamic fields as a function of the fluctuating metric tensor $\delta g_\mn$ and the 2-form gauge field~$\delta b_\mn$. One can then read out the correlators by inserting these expressions into the constitutive relations for $T^\mn$ and $J^\mn$ \eqref{eq:ConstRel} and taking functional derivatives, evaluated in equilibrium:
\begin{alignat}{2}
    G_{TT}^{\mn,\ab} &= -2 \frac{\delta T^\mn}{\delta h_\ab}, \qquad
    G_{JJ}^{\mn,\ab} &&= - \frac{\delta J^\mn}{\delta b_\ab},\label{eq:corrsDef1}\\
    G_{JT}^{\mn,\ab} &= -2 \frac{\delta J^\mn}{\delta h_\ab}, \qquad
    G_{TJ}^{\mn,\ab} &&= - \frac{\delta T^\mn}{\delta b_\ab}\label{eq:corrsDef2}.
\end{alignat}

The mixed $\langle JT\rangle$ and $\langle TJ \rangle$ correlators offer insight into new coefficients associated with heat and magnetothermal transport. For the clearest interpretation, it is convenient to introduce the heat current, $Q^\mu$. Its form can be deduced by reformulating the first law of thermodynamics covariantly:
\begin{equation}
    T S^\mu = p u^\mu - T^\mn u_\nu - \mu \, J^\mn h_\nu.
\end{equation}
This expression motivates the following definition:
\begin{equation}
    Q^\mu =- T^\mn u_\nu - \mu \, J^\mn h_\nu.
\end{equation}
In analogy to ordinary charged hydrodynamics, the heat current quantifies the difference between energy (momentum) flux and mechanical transport. We observe that the antisymmetry of $J^\mn$ ensures that no mechanical transport occurs along $h^\mu$.

We can now study correlators of the heat current with either $Q^i$, $J^\mn$ or $T^\mn$. Collectively denoting them as $X^I$, we have:
\begin{subequations}
\begin{align}
    G_{QX}^{\mu,I} &= -u^\eq_\nu G^{\mu\nu,I}_{TX}-\mu_\eq h^\eq_\nu G^{\mu\nu,I}_{JX},\\
    G_{XQ}^{I,k}&=-u^\eq_\alpha G_{XT}^{I,\alpha k} - \mu_\eq  h^\eq_\alpha G^{I,\alpha k}_{XJ},
\end{align}
\end{subequations}
which in turn yield
\begin{align}
    G_{QQ}^{i,k} = u_\nu u_\alpha G^{i\nu, \alpha k}_{TT} + \mu_\eq h_\alpha u_\nu\left(  G^{i\nu,\alpha k}_{TJ}+G^{i\alpha,\nu k}_{JT} \right) + \mu_\eq^2 h_\nu h_\alpha G^{i\nu,\alpha k}_{JJ}.
\end{align}

\subsubsection{Thermomagnetic transport coefficients}

To study thermomagnetic transport we focus on the spatial parts of the heat current $Q^i$ and the conserved 2-form current $J^{ij}$. These couple to the temperature gradient and $H_{tij}$ components of the gauge field curvature:
\begin{subequations}
\begin{align}
    & \delta J^{ij} = r^{ij,kl} H_{tkl} -\alpha^{ij,k}\partial_k  T,\\
    & \delta Q^i =T_\eq\tilde{\alpha}^{i,kl} H_{tkl} -  \bar{\kappa}^{i,k} \partial_k T.
\end{align}
\end{subequations}
With standard methods (which we outline in \Cref{app:thermo-derivation}), one can rewrite the sources in terms of external metric and gauge field perturbations and collect these linear relations in a compact form, schematically resembling the thermoelectric transport matrix in the case of ordinary charged hydrodynamics (see e.g. \cite{Hartnoll:2016apf}):
\begin{align}\label{eq:ThermoMagnetic_Matrix}
    \begin{pmatrix}
        \delta J^{ij}\\
        \delta Q^i
    \end{pmatrix} = \begin{pmatrix}
        r^{ij,kl} & T_\eq\alpha^{ij,k} \\
        T_\eq\tilde{\alpha}^{i,kl} & T_\eq\bar{\kappa}^{i,k}
    \end{pmatrix}\begin{pmatrix}
        i\omega \left[\delta b_{kl}+\mu_\eq\left(\delta_k^z\delta g_{tl}- \delta_l^z\delta g_{tk}\right)\right]\\
        i\omega \delta g_{tk}
    \end{pmatrix}.
\end{align}
Analogously to the thermoelectric effect, we call the object connecting the system response (LHS) with the sources on the RHS the \textit{thermomagnetic matrix} with the components of this matrix being the \textit{thermomagnetic transport coefficients}. Physically, $r^{ij,kl}$ and $\bar{\kappa}^{i,j}$ represent the electrical resistivity (which is usually, but not always, the inverse of the electrical conductivity \cite{Frangi:2024mer}) and the thermal conductivity (in presence of a non-zero current $J^{ij})$, respectively. The off-diagonal coefficients describe the von Ettinghausen effect, in which an external electric field applied in presence of a magnetic field creates heat flow. This should be roughly thought as the inverse of the more familiar Nernst effect (cf. \cite{Hartnoll:2007ih}).

Using the above relations \eqref{eq:ThermoMagnetic_Matrix}, one can read out the thermomagnetic transport coefficients as:
\begin{subequations}
\label{eq:ThermoMag}
\begin{align}\label{eq:ThermoMag_R}
    & r^{ij,kl}= \frac{1}{-i\omega  }\lim_{\mathbf{k}\rightarrow 0}\left( G_{JJ}^{ij,kl}(\omega,\mathbf{k}) - G_{JJ}^{ij,kl} (0,\mathbf{k})\right),\\
    & \alpha^{ij,k} = \frac{1}{-i\omega T_\eq }\lim_{\mathbf{k}\rightarrow 0}\left( G_{JQ}^{ij,k}(\omega,\mathbf{k}) - G_{JQ}^{ij,k} (0,\mathbf{k})\right),\label{eq:ThermoMag_a}\\
    & \tilde\alpha^{i,jk} = \frac{1}{-i\omega T_\eq}\lim_{\mathbf{k}\rightarrow 0}\left( G_{QJ}^{i,jk}(\omega,\mathbf{k}) - G_{QJ}^{i,jk} (0,\mathbf{k})\right),\label{eq:ThermoMag_aTilde}\\
    & \bar{\kappa}^{i,j} = \frac{1}{-i\omega T_\eq}\lim_{\mathbf{k}\rightarrow 0}\left( G_{QQ}^{i,j}(\omega,\mathbf{k}) - G_{QQ}^{i,j} (0,\mathbf{k})\right).\label{eq:ThermoMag_kappa}
\end{align}
\end{subequations}
In \Cref{app:WardMHD} we show that these transport coefficients obey the expected constraints imposed by Ward identities. Furthermore, the off-diagonal components of the thermomagnetic matrix satisfy the Onsager reciprocal relations:
\begin{align}
    \alpha^{i,jk}=\tilde{\alpha}^{jk,i}.
\end{align}
The independent components of the thermomagnetic transport matrix are
\begin{subequations}
\label{eq:tmtrcfs}
\begin{align}
    &r^{iz,kl} = 2\left(r_\perp + \frac{i}{\omega} \frac{\rho_0^2}{\varepsilon_0+p_0}\right)\eta^{i[k}h_\eq^{l]},\label{eq:ri3kl}\\
    &r^{xy,xy}=r_\parallel,\label{eq:r_FullTheoryII}\\
    &\alpha^{ij,k} = 2\left(\frac{\mu_\eq}{T_\eq}r_\perp - \frac{i}{\omega}\frac{s_0 \rho_0}{\varepsilon_0+p_0}\right)h_\eq^{[i}\eta^{j]k}, \label{eq:tmend} \\
    &\bar{\kappa}^{i,k}=\left(\frac{\mu_\eq^2}{T_\eq}r_\perp + \frac{i}{\omega}\frac{s_0^2 T_\eq}{\varepsilon_0+p_0}\right)\left(\eta^{ik}-h_\eq^i h_\eq^k\right) + \frac{i}{\omega}s_0 h_\eq^i h_\eq^k.
\end{align}
\end{subequations}
The Drude weight is given by the same thermodynamic expression as in ordinary hydrodynamics \cite{Hartnoll:2016apf} (also derived from microscopic descriptions such as kinetic theory \cite{Bajec:2024jez} and holography, see e.g.~\cite{Donos:2014cya}), with the main difference here being the more complicated index structure. Curiously, the parallel component of resistivity, $r^{xy,xy}$, does not exhibit the characteristic DC singularity in MHD. Furthermore, heat transport in the longitudinal direction, $\bar{\kappa}^{z,z}$, exhibits only the Drude peak without any finite DC part. This is a direct consequence of the antisymmetry of the conserved current $J^\mn$ and the Ward identities, as noted in \Cref{app:WardMHD}.

\subsection{Full linearized theory in canonical approach}\label{sec:full_MHD_Can}
In addition to the method outlined above, one can alternatively use the canonical approach to calculate the retarded correlators, which removes the need to use external sources. Even though the approaches are equivalent and result in identical retarded correlators (modulo contact terms), they provide complementary perspectives on the probe limit taken later. Furthermore, since the canonical approach to MHD has not, to our knowledge, been discussed in the literature, we describe it here in detail. The discussion here follows and generalizes \cite{Kovtun:2012rj} (see also \cite{Kadanoff:1963axw}).

We denote the seven conserved fluctuating charge densities as $\delta T^{tt} \equiv \delta \varepsilon$, $\delta T^{ti}\equiv \delta \pi^i$, $\delta J^{tx} \equiv \delta \rho^x$, $\delta J^{ty} \equiv \delta \rho^y$ and $\delta J^{tz} \equiv \delta \rho$, and arrange them into a tuple $\delta \varphi^a = (\delta\varepsilon,\delta \pi^i,\delta\rho^\perp,\delta\rho)$, where $\perp\,\in\{x,y\}$. Next, we identify the sources and susceptibilities of these charges. The sources are closely related to the fluctuating macroscopic fields $\delta T$, $\delta \mu$, $\delta u^i$ and $\delta h^i$. Taking the background \eqref{eq:hveq} implies that only $\delta h^x$ and $\delta h^y$ fluctuate (due to orthogonality). In a natural way, these present a candidate source for $\delta \rho^\perp$. We study this in more detail here.

The grand canonical density matrix describing our chosen background state is
\begin{align}
    & \varrho \sim \exp\spars{-\beta (H-\mu Q)},
\end{align}
where we neglect the normalization, which is not consequential for the sake of our argument. Locally, we can write (cf. \cite{Becattini:2014yxa})
\begin{align}
-\beta (H-\mu Q) &= -\beta \int_\Sigma d^3\Sigma_\mu T^\mn u_\nu +\beta\mu\int_{\tilde{\Sigma}} d^2\tilde{\Sigma}_\mn (\star J)^\mn ,\nonumber \\
& =-\beta \int_\Sigma d^3\Sigma\, u_\mu T^\mn u_\nu -\beta\mu\int_{\tilde{\Sigma}} d^2\tilde{\Sigma}\, u_\mu h_\nu J^\mn \label{eq:DensityMatrix1},
\end{align}
for spacelike surfaces $\Sigma$ and $\tilde{\Sigma}$ with codimensions $1$ and $2$, respectively. In the second line we used our only timelike and spacelike vectors in the theory to replace $d\Sigma_\mu$ and  $d\tilde{\Sigma}_\mn$ with $d\Sigma\,u_\mu$ and $\frac{1}{2}d\tilde{\Sigma}\,\varepsilon_{\ab\mn}u^\beta h^\alpha $, respectively. Perturbing the macroscopic fields $T,\mu,u^\mu$ and $h^\mu$ in \eqref{eq:DensityMatrix1}, we find
\begin{align}
   \beta_\eq \bigg[ \int d^3\Sigma\,\bigg( \frac{\delta T}{T_\eq} T^{tt}  +  \delta u_i T^{ti} \bigg) + \int dxdy\left(\delta \mu - \frac{\mu_\eq}{T_\eq}\delta T\right)J^{tz} + \int d^2\tilde{\Sigma}\,\mu_\eq\delta h_\perp J^{t\perp}\bigg],
\end{align}
from which we deduce the following charge-source relations
\begin{equation}
    \delta\varepsilon \leftrightarrow \frac{\delta T}{T_\eq}, \quad 
    \delta \pi^i \leftrightarrow \delta u^i, \quad \delta\rho \leftrightarrow \delta\mu - \frac{\mu_\eq}{T_\eq}\delta T, \quad
    \delta \rho^\perp \leftrightarrow \mu_\eq \delta h^\perp.
\end{equation}
We deduce the susceptibilities of $\delta \rho^\perp$ and $\delta \pi^i$ from the constitutive relation for $J^\mn$. It can be easily seen that at $\mathcal{O}(\partial^0)$,
\begin{align}
    \delta J^{0\perp} = \delta \rho^\perp = \rho_0\, \delta h^\perp,
\end{align}
from which we identify the corresponding susceptibility
\begin{align}
    \chi_{\rho^\perp} = \frac{\partial \rho^\perp}{\partial(\mu_\eq h^{\perp}) } = \frac{\rho_0}{\mu_\eq},
\end{align}
where the non-diagonal derivatives vanish, i.e., $\partial \rho^x/\partial h^y = 0$. Similarly, one can find
\begin{align}
    &\chi_{\pi^\perp} = \varepsilon_0+p_0,\\
    &\chi_{\pi^z}=\varepsilon_0+p_0-\mu_\eq\rr_0.
\end{align}
The susceptibility matrix relating the sources $(\delta T/T_\eq, \delta \mu - \mu_\eq \delta T/T_\eq)$ to charges $(\delta \varepsilon,\delta \rho)$ has off-diagonal components and is completely analogous to ordinary charged hydrodynamics:
\begin{align}\label{eq:energy_charge_susc}
    \begin{pmatrix}
        \delta \varepsilon\\[4pt]
        \delta \rho
    \end{pmatrix}=\begin{pmatrix}
       T_\eq \left(\frac{\partial \varepsilon}{\partial T}\right)_\mu + \mu_\eq \left(\frac{\partial \varepsilon}{\partial \mu}\right)_T & \;\;\left(\frac{\partial \varepsilon}{\partial \mu}\right)_T \\[10pt]
       T_\eq \left(\frac{\partial \rho}{\partial T}\right)_\mu + \mu_\eq \left(\frac{\partial \rho}{\partial \mu}\right)_T & \;\;\left(\frac{\partial \rho}{\partial \mu}\right)_T
    \end{pmatrix}\begin{pmatrix}
    \delta T/T_\eq\\[5pt]
    \delta \mu - \frac{\mu_\eq}{T_\eq} \delta T
    \end{pmatrix}.
\end{align}
The existence of off-diagonal components in the susceptibility matrix leads us to the conclusion that, as in ordinary charged hydrodynamics, suppressing $\delta T$ and $\delta u^\mu$ may not be possible in MHD with $\mu_0\neq0$ --- that is, in presence of a background magnetic field. 

To compactly express the charge-source relations, we collect the former in another tuple $\lambda^a=(\delta T/T_\eq, \delta u^i, \mu_\eq\delta h^\perp, \delta\mu - \mu_\eq \delta T/T_\eq)$ and the susceptibilities in a matrix $\chi^{a}{}_b$, thus obtaining:
\begin{align}
    \delta \varphi^a = \chi^{a}{}_b \lambda^b.
\end{align}
This allows us to write the Fourier-space conservation equations \eqref{eq:conservation} as
\begin{align}
    \left(-i\omega\, \delta^{a}_{b} + M^{a}{}_{b}(\omega,\bf{k})\right)\,\delta\varphi^b\equiv K^{a}{}_{b} \,\delta\varphi^b = 0,
\end{align}
from which retarded correlators are simply read out as
\begin{align}\label{eq:CanonicalCorrs_Definition}
    G^{\text{(can)}}_{ab}=-\left(\delta_{a}^{c}+i\omega \left(K^{-1}\right)_{a}{}^{c}\right)\chi_{c}{}_{b}.
\end{align}

To study the linearized hydrodynamic equations, it is convenient to factorize the charges into two subsectors, defined by their relation with the spatial momentum $k^i = (q,0,k)$. To avoid confusion with the rest of this work, we will refer to these subsectors as the $\mathbf k$-transverse and $\mathbf k$-longitudinal sectors.

\paragraph{k-transverse sector}--- The conservation equations for $\delta \pi^y$ and $\delta \rho^y$ constitute a self-consistent closed system of dynamical equations. Because $k^y=0$, we call this the $\mathbf k$-transverse sector. The linearized equations are:
\begin{subequations}
\begin{align}
    &-i\omega \,\delta \pi^y + \frac{1}{\varepsilon_0+p_0}\left(\eta_\perp q^2 + \eta_\parallel k^2\right)\delta \pi^y - ik\mu_\eq \delta \rho^y =0,\\
    &-i\omega \,\delta \rho^y + \frac{\mu_\eq}{\rho_0}\left(r_\parallel q^2 + r_\perp k^2\right)\delta \rho^y - \frac{ik}{\varepsilon_0+p_0} \delta \pi^y =0,
\end{align}
\end{subequations}
which we can compactly write as:
\begin{align}
    \left(K_\text{t}\right)^{a}{}_{b}\left(\delta \varphi_\text{t}\right)^b=\begin{pmatrix}
    -i\omega + k_i\gamma^{ij}_\eta k_j &-i\mu_\eq k\\
    -i\frac{\rho_0}{\varepsilon_0+p_0}k & -i\omega + k_i \gamma^{ij}_rk_i 
    \end{pmatrix}\begin{pmatrix}
        \delta \pi^y\\
        \delta \rho^y
    \end{pmatrix}=\begin{pmatrix}
        0\\
        0
\end{pmatrix},\label{eq:CompletelyTransvserseSystem}
\end{align}
with $\gamma_\eta$ and $\gamma_r$ were given in \eqref{eq:gammas}. The correlators are then expressed in matrix form as
\begin{align}
    (G_\text{t})^\text{(can)}_{ab} = - (\chi_\text{t})_{ab}-i\omega \left(K^{-1}_\text{t}\right)_{ac}(\chi_\text{t})^{c}{}_{b},
\end{align}
with $(\chi_\text{t})_{ab}=\text{diag}\,(\varepsilon_0+p_0, \rho_0/\mu_\eq)$. The analytic structure of the correlators in this sector exhibits sound modes, with the dispersion relation given by the Alfvén waves \eqref{eq:Dispersion_Alfven}. The correlators themselves agree identically with the variational approach (modulo the contact term $\varepsilon_0$ in the case of $G_{\pi^y\pi^y}$).

\paragraph{k-longitudinal sector}--- The rest of the hydrodynamic variables can be organized in a 5-tuple
$(\delta \varphi_l)^a = (\delta\varepsilon,\delta\rho,\delta\pi^x,\delta\pi^z,\delta\rho^x)$, with their evolution described by a second decoupled, $\mathbf k$-longitudinal sector:
\begin{align}
    (-i\omega \delta^{a}_{b} & + (M_l)^{a}{}_{b})(\delta \varphi_l)^b = 0,
\end{align}
where we defined the matrix:
\begin{align}
&(M_l)^{a}{}_{b}=
\nonumber\\
&\left(
\begin{array}{ccccc}
 0  & 0 & i q & i k & 0 \\
 \alpha_1  r_\perp q^2& \alpha_2  r_\perp q^2  & \frac{i \rr_0}{w_0}q & 0 & -\frac{ \mu_\eq  r_\perp}{\rr_0}kq \\
 i \beta_1 q & i \beta_2 q & \frac{\eta_\parallel k^2+\left(\zeta_\perp +\eta_\perp\right) q^2}{w_0}  & \frac{\zeta_\times +\eta_\parallel}{s_0
   T_\eq}q^2 & -i \mu_\eq k \\
 i \left(\beta_1-\rho_0\left(\frac{\partial \mu}{\partial \varepsilon}\right)_\rho\right) k& i  \left(\beta_2-\mu_\eq-\rho_0 \left(\frac{\partial \mu}{\partial \rho}\right)_\varepsilon\right)k & \frac{\zeta_\times + \eta_\parallel }{w_0}k q & \frac{\zeta_\parallel k^2+\eta_\parallel q^2}{s_0 T_\eq} & -i \mu_\eq q \\
 -\alpha_1  r_\perp k q & -\alpha_2  r_\perp k q & -\frac{i  \rr_0}{w_0}k & 0 & \frac{ \mu_\eq r_\perp}{\rr_0}k^2  \\
\end{array}
\right).
\end{align}
In defining the matrix, we introduced $w_0 = \varepsilon_0+p_0$ to denote the equilibrium enthalpy, and the following shorthands:
\begin{align}
    & \beta_1 = \left(\frac{\partial p}{\partial \varepsilon}\right)_\rho,\quad 
    \alpha_1 =\left(\frac{\partial \mu}{\partial \varepsilon}\right)_\rho-\frac{\mu_\eq}{T_\eq}\left(\frac{\partial T}{\partial \varepsilon}\right)_\rho, \nonumber \\
    & \beta_2 = \left(\frac{\partial p}{\partial \rho}\right)_\varepsilon,\quad
    \alpha_2 =\left(\frac{\partial \mu}{\partial \rho}\right)_\varepsilon-\frac{\mu_\eq}{T_\eq}\left(\frac{\partial T}{\partial \rho}\right)_\varepsilon.
\end{align}
The $\mathbf k$-longitudinal susceptibility matrix is of the form
\begin{align}
    (\chi_l)^{a}{}_{b} = \begin{pmatrix}
        \chi_{\varepsilon\varepsilon} & \chi_{\varepsilon\rho}\\
        \chi_{\rho\varepsilon} & \chi_{\rho\rho}
    \end{pmatrix} \oplus \text{diag}\left(w_0,\,T_\eq s_0,\,\frac{\rho_0}{\mu_\eq}\right),
\end{align}
with the first term corresponding to the $(\delta\varepsilon,\delta\rho)$ susceptibility matrix from \eqref{eq:energy_charge_susc}. In analogy to \cite{Kovtun:2012rj}, one can show the following thermodynamic identities hold
\begin{subequations}
\begin{align}
    &\alpha_1\chi_{\varepsilon\varepsilon}+ \alpha_2\chi_{\rho\varepsilon}=0,\\
    &\alpha_1\chi_{\varepsilon\rho}+
    \alpha_2\chi_{\rho\rho}=1,\\
    &\beta_1\chi_{\varepsilon\varepsilon}+
    \beta_2\chi_{\rho\varepsilon}=w_0,\\
    &\beta_1\chi_{\varepsilon\rho}+
    \beta_2\chi_{\rho\rho}=\rho_0,
\end{align}
\end{subequations}
as well as $\chi_{\rho\varepsilon}=\chi_{\varepsilon\rho}$, the latter being the condition we obtained in \Cref{sec:full_MHD_Var} for $G_{TJ} = G_{JT}$. The correlators can be read off from the definition \eqref{eq:CanonicalCorrs_Definition} and exhibit magnetosonic sound poles with the dispersion relation given by \eqref{eq:Dispersion_Magnetosonic}.

\subsection{Probe limit}\label{sec:probe-mhd}

As we mentioned in the introduction, in the probe limit of hydrodynamics one suppresses $\delta T$ and $\delta u^\mu$ with the aim of retaining fluctuations of internal conserved currents (in our case $J^\mn$) only. This is the viewpoint assumed in recent works on the EFT of MHD \cite{Vardhan:2022wxz,Vardhan:2024qdi}, and many works in the astrophysics community, such as \cite{goldreich1992magnetic}. The aim of this section is to take a step back and study the consequences of imposing the constraint:
\begin{equation}
    \label{eq:pb23}
    \delta T, \delta u^\mu, [\delta g^\mn] = 0,
\end{equation}
on the calculation presented in the previous sections (the square brackets are a reminder that $\delta g^\mn$ appears only in a variational approach). 

The main results of this section is that in presence of background magnetic fields, the probe limit inevitably breaks the conservation of $T^\mn$, and is therefore justified in physical systems in which the fluctuations of energy and momentum are small. Suppressing the energy-momentum conservation \eqref{eq:Tmn_Conservation} then leads to a nontrivial solution matching the  diffusive collective modes found in the EFT approach \cite{Vardhan:2022wxz}. 

\subsubsection{Variational approach}

For readability's sake, we remind the reader that the equilibrium state we are perturbing is given by
\begin{equation*}
    \tag{\ref{eq:hveq}}
    T_\eq=\const., \quad \mu_\eq=\const., \quad u_\eq^\mu = \delta^\mu_t, \quad h^\mu_\eq = \delta^\mu_z,
\end{equation*}
together with the background metric $\eta_\mn$ and the 2-form gauge field source $b_{\eq}^\mn = \mu_\eq u_{\eq}^{[\mu} h_{\eq}^{\nu]}$. 

The three independent equations in \eqref{eq:Jmn_Conservation} entirely specify $\delta \mu$, $\delta h^x$ and $\delta h^y$, and by consequence, the retarded $\langle JJ \rangle$ correlators. In Fourier space, with $k^\mu = (\omega,q,0,k)$, they read:
\begin{subequations}
\label{eq:ProbeSol}
\begin{align}
    & \delta \mu = -2\rho_0r_\perp q\frac{k\,\delta b_{tx} - q\,\delta b_{tz} - \omega\,\delta b_{xz}}{-i\rho_0 \left(\frac{\partial \rho}{\partial \mu}\right)_T + r_\perp \rho_0 \,q^2+r_\perp \mu_\eq\left(\frac{\partial \rho}{\partial \mu}\right)_Tk^2},\label{eq:ProbeSol1}\\
    & \delta h^x =2\left(\frac{\partial \rho}{\partial \mu}\right)_T r_\perp k\frac{k \, \delta b_{tx} - q\, \delta b_{tz}-\omega\,\delta b_{xz}}{-i\rho_0 \left(\frac{\partial \rho}{\partial \mu}\right)_T + r_\perp \rho_0 \,q^2+r_\perp \mu_\eq\left(\frac{\partial \rho}{\partial \mu}\right)_T k^2},\label{eq:ProbeSol2}\\
    & \delta h^y =2 \frac{r_\parallel q^2\,\delta b_{ty} + r_\perp k^2\, \delta b_{ty} + r_\parallel q\omega\, \delta b_{xy} - r_\perp \omega k \, \delta b_{yz}}{-i\rho_0\omega + r_\perp \mu_\eq k^2+r_\parallel \mu_\eq q^2}.\label{eq:ProbeSol3}
\end{align}
\end{subequations}
Plugging these expressions into \eqref{eq:Tmn_Conservation} yields two independent equations:
\begin{subequations}
\begin{align}
    &\omega \delta b_{xz} + q\delta b_{tz} - k\delta b_{tx} =0,\label{eq:Probe_constr1}\\
    &r_\parallel\mu_\eq  q\left( k \delta b_{xy} + q \delta b_{yz}\right) + i \rho_0 \left( k \delta b_{ty} - \omega  \delta b_{yz} \right) =0.
\end{align}
\end{subequations}
Enforcing these equations restricts the values which $\delta b_\mn$ can take, as they impose the following three conditions:
\begin{align}
    H_{txz}=0, \quad H_{tyz} = 0, \quad  H_{xyz} = 0 \;(\text{or } r_\parallel = 0).
\end{align}
Either choice for the third constraint leads, via \eqref{eq:ProbeSol}, to the vanishing of \textit{all} linear perturbations:
\begin{equation}
    \delta \mu =0,\; \delta h^x=0, \; \delta h^y=0 \quad\Rightarrow\quad J^\mn = J^\mn_\eq + \mathcal{O}(\delta^2).
\end{equation}
Because the only way to preserve energy-momentum conservation in the probe limit is to kill all hydrodynamic fluctuations, we can conclude that the two are strictly speaking incompatible. In particular, the momentum density in the two transverse directions ($x$ and $y$) is not conserved. Even so, \eqref{eq:ProbeSol} are legitimate solutions to the conservation equations of $J^\mn$ and can be used to retrieve retarded correlators within this approximation scheme (via \eqref{eq:corrsDef1}). From their analytic structure, we find that hydrodynamic modes must have one of the two following dispersion relations:
\begin{align}
    \label{eq:pbmd2}
    \omega_1 = -i \frac{\mu_\eq}{\rho_0}\left(r_\perp k^2+r_\parallel q^2 \right), \quad \omega_2 = - i r_\perp \pars{ \frac{ \mu_\eq}{\rho_0}k^2 + \left(\frac{\partial \rho}{\partial \mu}\right)_T^{-1} q^2 }.
\end{align}
After taking into account the different nomenclatures and discrete symmetries, these are exactly the dispersion relations obtained from the probe limit EFT (see Eq.~(26) of \cite{Vardhan:2022wxz}).\footnote{Specifically, the Kubo formulae ensure that $r_\parallel \leftrightarrow \sigma^\parallel$ and $r_\perp \leftrightarrow \sigma^\perp_1$, and by definition one has $\partial \rho/\partial \mu \leftrightarrow \chi_\parallel$ and $\rho_0/\mu_\eq \leftrightarrow \chi_\perp$.}

\subsubsection{Canonical approach}

In the canonical framework perturbations are not induced by external fields and hence cannot be restricted. Rather, one must determine under which conditions, if any, the hydrodynamic equations themselves permit the decoupling of $J^\mn$ and $T^\mn$. In practice, this requires the correlation matrix $G_{ab}^{\text{(can)}}$ to be block-diagonal.

It is immediately clear from \eqref{eq:energy_charge_susc} and \eqref{eq:CanonicalCorrs_Definition} that $G_{ab}^{\text{(can)}}$ couples energy density fluctuations ($\delta\varepsilon$) with charge density ones ($\delta \rho$). This is in complete analogy with ordinary charged hydrodynamics and should not come as a surprise because the susceptibility matrices \eqref{eq:energy_charge_susc} in the two cases are identical. To further strengthen the argument, the $\mathbf k$-transverse sector \eqref{eq:CompletelyTransvserseSystem} is coupled through equilibrium thermodynamic quantities, which cannot vanish if we wish to preserve the nature of the perturbed equilibrium state. This complementary viewpoint leads us to the same conclusion we reached with the variational approach, i.e., the probe limit leads can only be valid within an approximation scheme.

To complete our discussion, we note that we can still formally take the probe limit by ignoring the energy-momentum sector, just like in the previous section. In this case, we only perturb $\delta h^x,\delta h^y$ and $\delta \rho$ and use the canonical approach described in \Cref{sec:full_MHD_Can} to extract the probe limit correlators. Since these are identical to the ones computed using the variational approach, we do not rewrite them here.

\paragraph{Transport coefficients}--- We conclude by listing the transport coefficients. In the probe limit there is no $\langle TJ \rangle$ correlator by definition, so that the only transport coefficients are the $r^{ij,kl}$. Explicitly, they are:
\begin{subequations}
\label{eq:visc=res}
\begin{align}
    r^{iz,kl}_\text{probe}&= 2r_\perp \eta^{i[k}h^{l]}_0,\\
    r^{xy,xy}_\text{probe}&=r_\parallel,
\end{align}
\end{subequations}
which is identical to the DC finite part of the full theory result \eqref{eq:tmtrcfs}.

\section{Holography in the bulk probe limit}
\label{sec:grav_probe}

In the previous section, the adoption of the EFT toolkit of hydrodynamics led us to a few conclusions regarding the coupling between energy and magnetic flux transport in the IR regime. The rest of this work is dedicated to testing those conclusions through a concrete realization of a microscopic theory showcasing the hydrodynamic behavior discussed above. Our method of choice is the AdS/CFT correspondence \cite{Maldacena:1997re}, through which we construct a classical gravitational dual of a supersymmetric, strongly interacting QFT with the desired symmetry properties (discussed below).  

In the context of gravity, the probe limit is defined by requiring fluctuations of any matter fields to not backreact on the metric tensor. This is tantamount to studying classical fields on top of a fixed geometry, without any additional constraints that might arise from the Einstein equations. For this section only, we then assume that:
\begin{equation}
    \label{eq:gravpb}
    \delta g_\ind{MN} = 0,
\end{equation}
where uppercase Latin letters denote spacetime indices in the higher-dimensional dual theory. We will relax this assumption in the next section. In this restricted context, there are several techniques which allow a direct comparison between holographic calculations and the closed-time-path effective field theory formulations \cite{Grozdanov:2013dba,Crossley:2015evo,Haehl:2018lcu,Liu:2018kfw}. One of these, developed in \cite{deBoer:2015ija}, is based on sewing suitable Euclidean and Lorentzian spacetime patches together (as prescribed in \cite{Skenderis:2008dh, Skenderis:2008dg}), explicitly constructing the real-time fold complex contour upon which the EFT formalism is based. In this work, we instead adopt the approach proposed by Glorioso, Crossley and Liu in \cite{Glorioso:2018mmw}.\footnote{For a recent development on the topic, see \cite{Ammon:2025vod}.} As we review later, this is based on the complexification of the radial coordinate in the gravitational theory, which allows to consider a smooth path connecting the two AdS boundaries of an eternal black hole. The advantage offered by this technique is its capability to straightforwardly capture spatial derivatives, allowing a direct comparison with actions of the type discussed above.

\paragraph{Note}--- In this section, the notions of magnetic viscosity and electrical resistivity coincide. We adopt the notation $r_\parallel,~r_\perp$ and $r_H$ for both.

\subsection{Action and general considerations}

The purpose of this section is to analyze the hydrodynamic action derived by taking the gravitational probe limit \eqref{eq:gravpb}, and to compare it with the results obtained \cite{Vardhan:2022wxz,Vardhan:2024qdi} and in \Cref{sec:mhd}. To this end, the holographic dictionary ensures that we need to consider a theory of a 2-form gauge field in a curved geometry. The minimal such choice, introduced for the purposes of constructing a holographic dual to MHD in \cite{Grozdanov:2017kyl,Hofman:2017vwr} (see also \cite{Grozdanov:2018fic,DeWolfe:2020uzb}), is:
\begin{equation}
    \label{eq:full_action}
    S_\text{MHD}  = - \frac{1}{12}  \int_\mathcal M d^5 x \sqrt{-g} H_\ind{ABC} H^\ind{ABC},
\end{equation}
where $H=dB$ is the tensor strength of a 2-form gauge field $B$, which enjoys a $B\to B+d\Lambda$ gauge freedom (with $\Lambda$ an arbitrary 1-form). In principle, one could add a second, Chern-Simons-like quadratic term:
\begin{equation}
       S_\text{CS} =  \sigma \int_\mathcal M d^5 x \, B \wedge H,
\end{equation}
where $\sigma$ is a constant. This is easily proven to be a topological term. Applying distribution properties for wedge products and exterior derivatives (see e.g. \cite{Nakahara:2003nw}), one finds:
\begin{equation}
    \int_\mathcal M B \wedge dB = \frac{1}{2} \int_\mathcal M d\pars{B\wedge B} = \frac{1}{2} \int_\mathcal{\partial M} B\wedge B.
\end{equation}
Its topological nature implies that this term is invisible to the bulk equations of motion. Whether to consider values of $\sigma\neq0$ is a choice we can make at the end, when we reconstruct the boundary action. The equations of motion derived \eqref{eq:full_action} are then
\begin{equation}\label{eq:very_general_eom}
    \partial_\ind{A} \pars{\sqrt{-g} H^\ind{AMN}}=0.
\end{equation}
Ultimately, our goal is to calculate the transport coefficients of a microscopic theory defined on the equilibrium state \eqref{eq:hveq}. In the dual theory, this can be achieved by a Hodge-dualized version of the magnetic black brane solutions introduced in \cite{DHoker:2009mmn, DHoker:2009ixq}. For computational convenience, we find it however simpler to consider the slightly more general metric ansatz:
\begin{equation}
    \label{eq:generalised_metric}
    ds^2 = -f(r)dv^2 + 2 dv dr + \lambda_{ij}(r) dx^i dx^j,
\end{equation}
where $v$ is the ingoing Eddington-Finkelstein null coordinate, the spatial sector is taken to be diagonal (so that $\lambda_{ij} = \lambda_i \delta_{ij}$), and thermality of the background state is ensured by requiring $f(r)$ to have a simple zero at some radial position $r_h$. The metric is, of course, a consistent background solution of the 2-form field coupled to dynamical gravity.

To continue, we separate \eqref{eq:very_general_eom} into: 
\begin{subequations}
\begin{align}
    \label{eq:eoms1}
    & \partial_\nu\pars{\sqrt{-g} H^{r\mu\nu}} = 0, \\[4pt]
    \label{eq:eoms2}
    & \partial_\ind{A}\pars{\sqrt{-g} H^{\mu\nu \ind{A}}}  = 0.
\end{align}
\end{subequations}
As in the simpler case of diffusion, application of the Bianchi identity $dH=0$ makes enforcing the first equation unnecessary, simplifying the task at hand \cite{Glorioso:2018mmw}. A further simplification comes from adopting the radial gauge. Given that gauge transformations act on the 2-form field as $B\to B+d\Lambda$, one may set $B_{r\mu}=0$ by selecting:
\begin{equation}\label{eq:gauge_1form}
    \Lambda_\ind{A} = - \int_{r_c}^{r} dr' \, B_{r\ind{A}}(r',x).
\end{equation}
We notice that this transformation acts on the boundary value of the remaining components as:
\begin{equation}
    \label{eq:stsou}
    B_{\mu\nu}(\infty,x) \equiv b_{\mu\nu}(x) \to b_{\mu\nu}(x) + \partial_\mu A_{\nu}(x) - \partial_\nu A_{\mu}(x) = b_{\mu\nu}(x) + f_{\mu\nu}(x),
\end{equation}
where we defined $A_\mu(x) = \Lambda_\mu(r\to\infty,x)$. This identifies $A_\mu$ with the Stückelberg field appearing in the EFT of \cite{Grozdanov:2016tdf}. As it appears only in the $f$ combination, it enjoys an ordinary $U(1)$ gauge freedom. The gauge transformation specified by \eqref{eq:gauge_1form} can be seen as a map between bulk configurations and Stückelberg fields. When we perform functional integration on all Stückelberg configurations, we are doing so on an equivalence class of bulk fields (conjugated by a gauge transformation). Following the analogous discussion by Nickel and Son for the case of charge diffusion \cite{Nickel:2010pr}, $A_\mu$ may be identified with a Goldstone boson associated to the spontaneous breaking of a doubled $U(1)$ higher-form symmetry to its diagonal subgroup: $U(1)\times U(1) \to U(1)$. The original two copies live on the AdS boundary and on a stretched horizon.

\subsection{Solving the EOMs}

Using the metric \eqref{eq:generalised_metric}, its inverse and the radial gauge $B_{r\mu}=0$, \eqref{eq:eoms2} can be broken down into:
\begin{subequations}
\label{eq:chEOM}
\begin{align}
    \label{eq:atEOMtime}
    & \partial_r (\sqrt{\lambda}\lambda^i B_{vi,r} ) - \partial_j (\sqrt\lambda\lambda^i {\textstyle \sum\nolimits_{j}} \lambda^j B_{ij,r}) = 0, \\[4pt]
    \label{eq:atEOMspace}
    & \partial_r \Big[ \sqrt\lambda \lambda^i \lambda^j (f B_{ij,r} + H_{vij}) \Big]  +  \sqrt\lambda \lambda^i \lambda^j ( B_{ij,rv} +  {\textstyle \sum\nolimits_{k}} \lambda^k H_{ijk,k} ) = 0.
\end{align}
\end{subequations}
We will refer to \eqref{eq:atEOMtime} as the \textit{temporal} equation of motion, and \eqref{eq:atEOMspace} as the \textit{spatial} one. Sums are kept explicit to avoid ambiguities. To solve these equations, we need to specify both a domain -- corresponding to the manifold $\mathcal M$ upon which the action \eqref{eq:full_action} is defined -- and boundary conditions. We start from the former.

\paragraph{On the domain of the EOMs}--- According to the real-time holographic prescription for computing thermal Green's function \cite{Son:2002sd,Herzog:2002pc}, the equations of motion need to be solved in the exterior of an eternal black hole solution -- that is, the radial domain considered in $\mathcal M$ is the infinite interval $(r_h,\infty)$. This is the region of the spacetime that can be `analytically continued' to the corresponding cigar-shaped Euclidean black hole. Boundary conditions are consequently imposed at some cutoff brane close to the AdS boundary (so to allow for renormalization) and on a stretched event horizon. This is not the case in our closed-time-path construction, in which we work with $\partial \mathcal M$ equalling two copies of the boundary theory, consistently with the Schwinger-Keldysh formalism.

\begin{figure}
	\centering
	\begin{tikzpicture}[scale=1]
		\node[rectangle,minimum width=.5cm,minimum height=.5cm,anchor=north east,draw] (lab) at (8,2) {$r$};
        \node[anchor = east] (inf1) at (8,-.5) {$\infty_1$};
        \node[anchor = east] (inf2) at (8,.5) {$\infty_2$};
        \draw[very thin,gray] 
			(0,0) -- (8,0)
			(0,-2) -- (0,2);
        \draw [domain=30:330] plot[smooth] ({2+cos(\x)}, {sin(\x)});
        \draw
            ($(2,0)+cos(30)*(1,0)+sin(30)*(0,1)$) -- (inf2) node[currarrow,pos=0.5,xscale=1,sloped,scale=1.25] {};
        \draw
            ($(2,0)+cos(30)*(1,0)-sin(30)*(0,1)$) -- (inf1) node[currarrow,pos=0.5,xscale=-1,sloped,scale=1.25] {};
		\fill [red] (2,0) circle (3pt);
        \node[inner sep=0pt,anchor=north] at (2,-0.2) {$r_h$};
        \fill [blue] ($(2,0)+cos(135)*(1,0)+sin(135)*(0,1)$) circle (3pt);
        \node[inner sep=0pt,anchor=south east] at ($(2,0)+1.2*cos(135)*(1,0)+1.2*sin(135)*(0,1)$) {$r_c$};
        \end{tikzpicture}
		\caption{The complex radial contour connecting the two AdS boundaries of an eternal black hole, upon which the EOMs \eqref{eq:chEOM} are defined. $r_c$ is the stretched horizon upon which the extra boundary condition \eqref{eq:dss_condition} is imposed.}
	\label{fig:contour}
\end{figure}
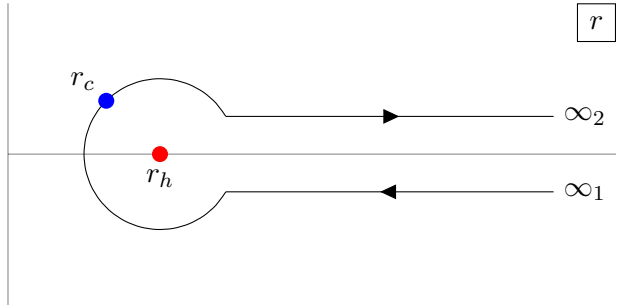

To understand this construction, one can identify points in $\mathcal M$ that are connected by a Killing flow and realize that the difference can be stated in term of radial curves upon which the EOMs are defined. The standard Son-Starinets prescription \cite{Son:2002sd} is defined on a curve connecting the AdS boundary to the event horizon. To connect the two AdS boundaries of an eternal black hole, a curve has to pass through its interior of the black hole. This can be achieved by complexifying the radial coordinate and consider the contour in \Cref{fig:contour}, as elucidated in a work by Festuccia and Liu \cite{Festuccia:2005pi}.  This is the radial domain we consider in this work.

\paragraph{Boundary conditions}--- When working in the radial gauge, boundary conditions to \eqref{eq:chEOM} include a combination of sources and Stückelberg fields, as seen in \eqref{eq:stsou}. At the two endpoints of the keyhole, we write them as:
\begin{equation}\label{eq:bcs_inf}
    B_{\mu\nu}\pars{r\to\infty_s,x} = b_{(s)\mu\nu}(x) + f_{(s)\mu\nu}(x) \equiv G_{(s)\mu\nu}(x), \quad s=1,2. 
\end{equation}
The $G$-fields will be the building blocks for our EFT. Following \cite{Vardhan:2022wxz}, we use them to define appropriate Keldysh variables:
\begin{equation}
	G_{(r)\mu\nu} = \frac{1}{2} \pars{G_{(1)\mu\nu} + G_{(2)\mu\nu}}, \quad G_{(a)\mu\nu} = G_{(1)\mu\nu} - G_{(2)\mu\nu}.
\end{equation}
After imposing the radial gauge condition $B_{r\mu}=0$, the residual gauge symmetry group is given by:
\begin{equation}
    B_{\mu\nu}(r,x) \to B_{\mu\nu}(r,x) + (d\Lambda_{(s)})_{\mu\nu}(x),
\end{equation}
with $\Lambda$ a 1-form independent of $r$. These transformations act nontrivially on the boundary conditions \eqref{eq:bcs_inf} to give:
\begin{align}
    \label{eq:residual_bdygauge}
        & G_{(r)\mu\nu}(x) \to G_{(r)\mu\nu}(x) + \frac{1}{2}(d\Lambda_{(1)} + d\Lambda_{(2)})_{\mu\nu}(x), \nonumber \\[4pt]
        & G_{(a)\mu\nu}(x) \to G_{(a)\mu\nu}(x) + (d\Lambda_{(1)} - d\Lambda_{(2)})_{\mu\nu}(x).
\end{align}

This is too large of a symmetry group. In absence of superfluid condensation we know that, besides spacetime translations, the only additional symmetry transformations are diagonal shifts given by:
\begin{equation}
	G_{(r)ij}(\mathbf x) \to G_{(r)ij}(\mathbf x) + d\Lambda_{(r)ij}(\mathbf x).
\end{equation}
Thus, we need to further restrict to the `spatial part' of the diagonal subgroup of the original higher-form $U(1) \times U(1)$ gauge symmetry. This can be achieved by imposing an extra boundary condition on the time components of $B$, on a stretched horizon $r_c = r_h + \varepsilon$ (where in general $\varepsilon \in \mathbb C$):
\begin{equation}\label{eq:dss_condition}
    B_{vi}\pars{r=r_c,x} = f(x).
\end{equation}
Throughout this work, we use $f=0$. This is effectively a gauge choice, and any nontrivial profile for $f$ can easily be reabsorbed by a suitable transformation.

Interestingly, \eqref{eq:dss_condition} formally breaks the integration domain of the time equation of motion into two intervals $(\infty_1,r_c)$ and $(r_c,\infty_2)$, in a way that looks very similar to the regularity condition that is usually imposed in the Son-Starinets prescription. In fact, while the condition ensures continuity of $B_{0i}$ across the whole keyhole, its radial derivative may be discontinuous.

\paragraph{Derivative expansion}--- The next step is to expand the equations of motion \eqref{eq:chEOM} in the number of boundary derivatives and solve them order by order.\footnote{We remind the reader that at the UV boundary, $\partial_v=\partial_t$.} This is done by introducing a small expansion parameter $\epsilon$:
\begin{equation}\label{eq:acoustic_scheme}
    (\partial_v,\partial_i) \to \epsilon (\partial_v,\partial_i), \quad B_{vi} \to \sum_{I=0}^{\infty}\epsilon^I B^{(I)}_{vi},\quad B_{ij} \to \sum_{I=0}^{\infty}\epsilon^I B^{(I)}_{ij}.
\end{equation}
Plugging this into \eqref{eq:chEOM} gives a tower of equations which can be solved order by order in $\epsilon$. All that is left to do is to assign a weight to the $G$-fields -- that is, to appropriately impose the boundary conditions at infinity \eqref{eq:bcs_inf}. In this work, we follow the original reference~\cite{Glorioso:2018mmw} and take:
\begin{gather}
    B^{(0)}_{\mu\nu}\pars{r\to\infty_s,x} = G_{(s)\mu\nu},\quad B^{(I)}_{\mu\nu}\pars{r\to\infty_s,x} = 0 \text{ for: }I\geq1, \nonumber \\[4pt] 
    \label{eq:quantum_BCs}
    B^{(I)}_{vi}\pars{r=r_c,x} = 0 \text{ for: } I \geq 0.
\end{gather}
Notice that if we are to assume that each $B_{\mu\nu}^{(n)}$ has a well-defined $\epsilon$-weight, then this choice of boundary conditions would be inconsistent with the classical KMS condition:
\begin{equation}\label{eq:classicalKMS}
    \varphi_a(x) \to \Theta \varphi_a( x) - i \beta_0 \Theta (\partial_v \varphi_r(x)).
\end{equation}
This is apparent in that it assigns the same weight to $r$ and $a$ variables, whereas from~\eqref{eq:classicalKMS} they should differ by one time derivative. A way to impose boundary conditions consistently with it would be:
\begin{align}
    & B^{(0)}_{\mu\nu}\pars{r\to\infty_s,x} = G_{(r)\mu\nu}(x),\quad B^{(1)}_{\mu\nu}(r\to\infty_s,x) = \frac{(-1)^{s+1}}{2} G_{(a)\mu\nu}(x), \\[4pt] & 
    \label{eq:classical_BCs}
    B^{(I)}_{\mu\nu}\pars{r\to\infty_s,x} = 0 \text{ for: }I \geq 2, \quad  B^{(I)}_{vi}\pars{r=r_c,x} = 0 \text{ for: } I \geq 0.
\end{align}
This is not the only ambiguity concerning the derivative expansion. One could also imagine assigning a double weight to time derivatives, which naturally appears in cases with diffusion:
\begin{equation}\label{eq:diffusive_dexpansion}
    (\partial_v,\partial_i) \to (\epsilon^2 \partial_v, \epsilon \partial_i).
\end{equation}
This would lead to a different counting scheme altogether. To account for this ambiguity, we carried out independent calculations for each of the 4 cases we mentioned and noticed that the overall effect to adopt a different scheme is to change the order at which a given term appears, leading to the same final result. This removes the necessity to consider each case separately.

\paragraph{Solution}--- Using the counting scheme in \eqref{eq:acoustic_scheme}, to first order in $\epsilon$, the equations of motion are:
\begin{itemize}
    \item Order 0:
        \begin{subequations}
        \begin{align} 
            \label{ac_eom:start}
            & \partial_r(\sqrt\lambda \lambda^i B^{(0)}_{vi,r}) = 0,\\[4pt]
            & \partial_r(\sqrt\lambda \lambda^i \lambda^j f B^{(0)}_{ij,r}) = 0.
        \end{align}
        \end{subequations}
    \item Order 1:
         \begin{subequations}
         \begin{align}
            & \partial_r(\sqrt\lambda \lambda^i B^{(1)}_{vi,r}) - \sqrt\lambda \lambda^i {\textstyle \sum\nolimits_j \lambda^j B^{(0)}_{ij,rj}}= 0, \\[4pt]
            & \partial_r\big[\sqrt\lambda \lambda^i \lambda^j (f B^{(1)}_{ij,r} + H^{(0)}_{vij})\big] + \sqrt\lambda \lambda^i \lambda^j B^{(0)}_{ij,rv}= 0.
        \end{align}
        \end{subequations}
\end{itemize}
The boundary conditions are given by \eqref{eq:quantum_BCs}. We solve them on the complex radial contour of \Cref{fig:contour}.

\paragraph{Order 0:} 
The leading order solutions are:
\begin{subequations}
\begin{align}
    & B_{vi}^{(0)}(r,x) = \bar G_{(s)ti}(x) b_i(r), \\[4pt]
    & B_{ij}^{(0)}(r,x) = G_{(r)ij}(x) - \bar G_{(a)ij}(x) l_{ij}(r),
\end{align}
\end{subequations}
where the following radial functions have been introduced:
\begin{equation}
    b_i(r) = \int_{r_c}^r dr'\,\frac{\lambda_i}{\sqrt\lambda},\quad \zeta_{ij}(r) = \int_{\infty_1}^r dr'\, \frac{\lambda_i \lambda_j}{\sqrt\lambda f}, \quad l_{ij}(r) = \zeta_{ij}(r) - \frac{1}{2}\kappa_{ij},
\end{equation}
together with the constants:
\begin{equation}
    \quad Q_i = b_i(\infty_s), \quad \kappa_{ij} = \zeta_{ij}(\infty_2),
\end{equation}
and the shorthand notation $\bar G$, defined as:
\begin{equation}
    \bar G_{(\ind{\bullet})ti}(x) = \frac{G_{(\ind{\bullet})ti}(x)}{Q_i}, \quad \bar G_{(\ind{\bullet})ij}(x) = \frac{G_{(\ind{\bullet})ij}(x)}{\kappa_{ij}}.
\end{equation}
Notice that the constant $Q_i$ is independent from which boundary we use to calculate it, as the integrand that defines it is holomorphic in the complex $r$ plane.

\paragraph{Order 1:}
At first order, the solutions are:
\begin{subequations}
\begin{align}
    & B_{vi}^{(1)}(r,x) = -  p_{(a)i}(x) \xi_i(r), \\[4pt]
    & B_{ij}^{(1)}(r,x) = \bar G_{(a)ij,t}(x) a(r) l_{ij}(r) - \bar G_{(s)ti,j}(x) \tilde b_{i}(r) + \bar G_{(s)tj,i}(x) \tilde b_{j}(r) - \nonumber \\ & \hspace{.5\textwidth} - G_{(r)ij,t}(x)a(r) + \tau_{ij}(x) \zeta_{ij}(r).
\end{align}
\end{subequations}
The newly introduced radial functions are:
\begin{equation}
    a(r) = \int_{\infty_1}^r \frac{dr'}{f}, \quad \xi_i (r) = \int_{r_c}^r dr'\, \frac{\lambda_i}{\sqrt\lambda}\pars{a(r') - \frac{R_i}{Q_i}}, \quad \tilde b_i (r) = \int_{\infty_1}^r dr'\, \frac{b_i}{f},
\end{equation}
and the constants:
\begin{equation}
    \kappa = a(\infty_2),\quad \Lambda_i = \tilde b_i(\infty_s), \quad R_i = \int_{r_c}^{\infty_s} dr \, \frac{\lambda_i}{\sqrt\lambda} a(r).
\end{equation}
For the same reason as $Q_i$, neither $\Lambda_i$ nor $R_i$ depend on the boundary we use to calculate them. We have also used:
\begin{align}
    & p_{(a)i}(x) = {\textstyle \sum}_{j} \bar G_{(a)ij,j}(x), \nonumber \\[4pt]
    & \tau_{ij}(x) = - \frac{1}{\kappa_{ij}}\pars{ \frac{1}{2} \kappa \kappa_{ij} \bar G_{(a)ij,t} + \Lambda_i \bar G_{(a)ti,j} - \Lambda_j \bar G_{(a)tj,i} - \kappa G_{(r)ij,t}},
\end{align}
where, to find the latter expression, we had to calculate integrals of the type:
\begin{equation}
    G_{(s)ti}(x)\oint dr\, h(r),
\end{equation}
with $h$ a holomorphic function. These are in general finite due to the non-analiticity introduced by $G_{(s)}$ terms. The proper way to handle these kind of integrals is to separate the keyhole in two branches with the explicit expression for $G_{(s)}$:
\begin{align}
    G_{(s)ti}(x)\oint dr\, h(r) &= \pars{G_{(r)ti}(x) + \frac{1}{2} G_{(a)ti}(x)}  \oint dr\, h(r) - G_{(a)ti}(x) \int_{r_c}^{\infty_2} dr\, h(r) \nonumber\\ &= - G_{(a)ti}(x) \int_{r_c}^{\infty_2} dr\, h(r).
\end{align}

\subsection{Holographic CTP effective action}

To find the effective boundary CTP action, we plug the solutions into the bulk action \eqref{eq:full_action} and solve the radial integral on the path indicated in \Cref{fig:contour}. The action retains the order of the derivative expansion in the bulk. Let us start from the non-topological sector of the action. Making use of the metric \eqref{eq:generalised_metric} and the radial gauge condition $B_{r\mu}=0$, we find:
\begin{multline}\label{eq:action_exp1}
    S_\text{MHD} = -\frac{1}{4} \int d^4 x \int_C dr \sqrt{\lambda} \Big[ \lambda^i \lambda^j B_{ij,r} \pars{ f B_{ij,r} + 2 B_{ij,v} + 4 B_{vi,j} } - \\ - 2 \lambda^i \pars{B_{vi,r}}^2 + \frac{1}{3} H_{ijk} H^{ijk}  \Big].
\end{multline}
A sum over $i,j,k$ is implied. Rather than substituting and calculating the integrals by brute force, we use a few properties of the effective action in order to distill its contents. First, we can use the derivative expansion to observe that, at first order in $\epsilon$, \eqref{eq:action_exp1} can be rewritten as:
\begin{multline}\label{eq:action_exp2}
    S_\text{MHD} = -\frac{1}{4} \int d^4 x \int_C dr \sqrt{\lambda} \Big[ f \lambda^i \lambda^j B^{(0)}_{ij,r} (B^{(0)}_{ij,r} + 2 B^{(1)}_{ij,r}) + 2 \lambda^i\lambda^j B^{(0)}_{ij,r}B^{(0)}_{ij,v} + \\ + 4 \lambda^i \lambda^j B^{(0)}_{ij,r}B^{(0)}_{vi,j} - 2 \lambda^i B^{(0)}_{vi,r} (B^{(0)}_{vi,r} + 2 B^{(1)}_{vi,r})
    \Big].
\end{multline}
We can now perform the radial integrals. These are very similar to the ones we performed in the previous section. For instance, one has:
\begin{equation}
    \int_C dr \sqrt{\lambda}  f \lambda^i \lambda^j (B^{(0)}_{ij,r})^2 = \kappa_{ij} \bar G_{(a)ij}^2 = \frac{1}{\kappa_{ij}} G_{(a)ij}^2. 
\end{equation}
Many of the integrals vanish because the integrands are regular, e.g.,
\begin{equation}
    \int_C dr \sqrt{\lambda}  f \lambda^i \lambda^j B^{(0)}_{ij,r} B^{(0)}_{ij,v} = \bar G_{(a)ij} \bar G_{(a)ij,t} \int_C \zeta_{ij} = 0. 
\end{equation}

All in all, this leads to the following action:
\begin{align}
    \label{eq:simple_action}
    \mathcal L_\text{MHD} = - \frac{1}{4 \kappa_{ij}} G_{(a)ij}(x)^2 + & \frac{1}{2 Q_i} G_{(r)ti}(x) G_{(a)ti}(x)  + \frac{\kappa}{2 \kappa_{ij}} G_{(r)ij,t}(x) G_{(a)ij}(x) +\nonumber \\  & \quad + \Gamma_{ij} G_{(a)ij,t}(x) G_{(a)ij}(x) - \frac{\Lambda_i}{\kappa_{ij} Q_i} G_{(a)ij}(x) G_{(a)ti,j}(x),
\end{align}
where we introduced a composite constant:
\begin{equation}
    \Gamma_{ij} = - \frac{\kappa}{4 \kappa_{ij}} + \frac{1}{2 \kappa_{ij}^2}\int_{\infty_1}^{\infty_2} dr \, \frac{\lambda_i \lambda_j}{\sqrt\lambda f} a.
\end{equation}
Repeating the above for the topological part leads to
\begin{align}
        \mathcal L_\text{CS} =  - 2 \sigma \varepsilon_{ijk} \Big( G_{(r)ij} G_{(a)tk} + G_{(a)ij} & G_{(r)tk} - 2 \frac{\Lambda_i}{Q_i} G_{(a)ti} G_{(r)tk,j} -  \frac{\Lambda_i}{Q_i} G_{(a)tk} G_{(a)ti,j} \Big) .
\end{align}
We observe that there are no $r-r$ terms in the effective Lagrangian -- as expected from the general arguments of \cite{Crossley:2015evo} -- and that only part of the coefficients need renormalization. Those that do not (which happen to be the ones we are interested in) are universal, as they only depend on some combination of metric coefficients evaluated at the horizon.

\subsection{Transport coefficients in presence of background magnetic fields}

So far, we solved the problem for the general diagonal metric ansatz \eqref{eq:generalised_metric}. As stated at the beginning of this section, though, we are interested in studying fluctuations on top of the equilibrium state specified by \eqref{eq:hveq}. This corresponds to taking a magnetic black brane background, which has the metric:
\begin{equation}
	\label{eq:genmb}
    ds^2 = -U(r)dv^2 + 2dvdr + e^{2V(r)}(dx^2+dy^2) + e^{2W(r)}dz^2.
\end{equation}
In terms of the coefficients used so far, it is simple to find, by comparing the expression with \eqref{eq:generalised_metric}:
\begin{equation}
    f = U, \quad \lambda_x = \lambda_y = e^{2V}, \quad \lambda_z = e^{2W}.
\end{equation}
Recalling the definitions of the coefficients in the effective Lagrangian:
\begin{equation}
    \label{eq:ceeld}
    Q_i = \int_{r_c}^{\infty_s} dr \, \frac{\lambda_i}{\sqrt\lambda}, \quad \kappa_{ij} = \int_{\infty_1}^{\infty_2} dr \, \frac{\lambda_i\lambda_j}{\sqrt\lambda f}, \quad \Lambda_i = \int_{r_c}^{\infty_s} \frac{dr}{f} \int_{r_c}^r dr' \, \frac{\lambda_i}{\sqrt\lambda},
\end{equation}
we find:
\begin{equation}
    Q_x = Q_y \neq Q_z, \quad \kappa_{xy} \neq \kappa_{xz} = \kappa_{yz}, \quad \Lambda_x = \Lambda_y \neq \Lambda_z.
\end{equation}
It follows that the $r-a$ part of the action (which, as we will see in a moment, is the only one contributing to the constitutive equations) becomes:
\begin{multline}
    \mathcal L^{(ra)} = \frac{\kappa}{\kappa_{xy}} G_{(r)xy,t} G_{(a)xy} + \frac{\kappa}{\kappa_{xz}} \sum_{a=x,y} G_{(r)az,t} G_{(a)az} + \frac{1}{2 Q_x} \sum_{a} G_{(r)ta}^2 + \frac{1}{2 Q_z} G_{(r)tz}^2 + \\ + 4 \sigma (G_{(a)xy} G_{(r)tz} - \sum_{a,b} \varepsilon_{ab} G_{(a)az} G_{(r)tb}) + 4 \sigma [ G_{(a)tz } (G_{(r)xy} - \frac{2 \Lambda_z}{Q_z} G_{(r)tx,y}) + \\ + \sum_{a,b} \varepsilon_{ab} G_{(a)tb} (G_{(r)az} - \frac{2 \Lambda_x}{Q_x} G_{(r)ta,z})].
\end{multline}
Then, the constitutive equations for the conserved 2-form current $J^{\mu\nu}$ are obtained by taking appropriate functional derivatives:
\begin{equation}
    \label{eq:def_VEV_J}
    J^{(r)\mu\nu} = \frac{\delta S_\text{MHD+CS}}{\delta G_{(a)\mu\nu}},\quad J^{(a)\mu\nu} = \frac{\delta S_\text{MHD+CS}}{\delta G_{(r)\mu\nu}}.
\end{equation}
It is manifest from the absence of $r-r$ terms in \eqref{eq:simple_action} that $G_{(a)\mu\nu}=0$ solves the latter equation. The conserved $r$-current (from which we drop the variable index) reads:
\begin{subequations}
\label{eq:ce_here}
\begin{align}
        & J^{ta} = \frac{1}{Q_x} G_{ta} - 4 \sigma \varepsilon_{ab} (G_{bz} - \frac{2 \Lambda_x}{Q_x} G_{tb,z}), \\[4pt]
        & J^{tz} = \frac{1}{Q_z} G_{tz} + 4 \sigma (G_{xy} - \frac{2 \Lambda_z}{Q_z} G_{tx,y}), \\[4pt]
	& J^{az} = \frac{\kappa}{\kappa_{xz}} G_{az,t} + 4 \sigma \varepsilon_{ab} G_{tb}, \\[4pt]
	& J^{xy} = \frac{\kappa}{\kappa_{xy}} G_{xy,t} + 4 \sigma G_{tz}.
\end{align}
\end{subequations}
Transport coefficients can be read off from the constitutive relations either by taking another functional derivative (this time in $G_{(r)}$) or by comparing them with the ones obtained in the probe limit EFT of \cite{Vardhan:2022wxz}. Such a comparison can be made after setting $\sigma=0$, in order to remove topological anomalous effects, giving us
\begin{equation}
    \label{eq:pb_tcs}
    \chi_\parallel = \frac{1}{Q_z}, \quad \chi_\perp = \frac{1}{Q_x}, \quad r_\parallel = \frac{\kappa}{\kappa_{xy}}, \quad r_\perp = \frac{\kappa}{\kappa_{xz}}, \quad r_H = 0.
\end{equation}
The explicit expressions of the EFT coefficients lead to the conclusion that the resistivities do not depend on the renormalization procedure. Because magnetic brane solutions are known only numerically, we delay the analysis of their behavior as functions of the thermodynamic parameters to the next section, where we outline the details of the numerical calculation.

\subsection{Summary and potential issues}

In this section, we used the holographic method proposed in \cite{Glorioso:2018mmw} to construct a hydrodynamic effective action for a conserved 2-form current, in the bulk probe limit in which the metric fluctuation are frozen, mirroring the structure of the probe limit in hydrodynamics. We then matched our constitutive equations with the ones obtained from the hydrodynamic probe limit EFT described in \cite{Vardhan:2022wxz}, and thus found explicit formulae for the two resistivities $r_\parallel,~r_\perp$ and the vanishing Hall resistivity $r_H = 0$.

This last point is particularly interesting: based on the symmetry principles elucidated in \cite{Vardhan:2022wxz,Vardhan:2024qdi}, one would expect $r_H\neq0$ whenever charge conjugation symmetry is broken. The existence of magnetic brane solutions with finite charge density (which breaks $C$) and a diagonal metric tensor \cite{DHoker:2009ixq}, which as we show below lead to $r_H\neq0$ in the full backreacting theory, suggests that a probe limit realization of $r_H$ could require the introduction of additional fields, such as a massive 1-form vector. This would amount to breaking charge conjugation symmetry with the operator content of the theory, rather than the choice of state, and would lead to a holographic realization of a probe limit EFT more similar to the one we expect to describe neutron stars. 

\section{Beyond the bulk probe limit}
\label{sec:beyond}

The aim of this section is to enable a comparison between the probe-limit results of the previous section with the calculations of transport coefficients from the full theory, using standard (non-EFT) holographic techniques. We consider the action coupling the 2-form gauge field to the Einstein-Hilbert gravity with a negative cosmological constant \cite{Grozdanov:2017kyl,Hofman:2017vwr}:
\begin{equation}
    \label{eq:uberS}
    S_\text{full} = \int d^5 x \sqrt{-g} \pars{ R + 12 - \frac{1}{12} H_\ind{ABC}H^\ind{ABC}}.
\end{equation}
We will also outline the steps necessary to numerically determine the neutral and (nonanomalous) charged magnetic brane backgrounds. In the process, we extend the earlier analysis of \cite{Grozdanov:2017kyl}.

The starting point is the covariant equations of motion:
\begin{subequations}
\begin{align}
    \label{eq:beom}
    & \partial_\ind{A}(\sqrt{-g} H^\ind{ABC})=0, \\
    \label{eq:meom}
    & R_\ind{AB} - \frac{1}{2}(R + 12) g_\ind{AB} - \frac{1}{4} \pars{ H_\ind{ACD}{H_\ind{B}}^\ind{CD} - \frac{1}{6}H^2 g_\ind{AB} } = 0.
\end{align}
\end{subequations}
For the background, we consider a diagonal magnetic brane ansatz. Adopting the timelike $t$ instead of the null ingoing $v$ of the previous section, this reads:
\begin{subequations}
    \begin{align}
        & H = e^{W-2V} \rho \, dt\wedge dz\wedge dr + n \, dx \wedge dy \wedge dz, \label{eq:hfceq} \\
        & ds^2 = - U(r)dt^2 + e^{2 V(r)} (dx^2 + dy^2) + e^{2W(r)} dz^2 + \frac{dr^2}{U(r)}. \label{eq:metr}
    \end{align}
\end{subequations}
The dimensionful constants, $\rho$ and $n$, correspond (up to a constant determined by a rescaling of the $\mathbf x$ coordinates, see later) to the background magnetic field (pointed along $z$) and charge density, respectively. The charge density gives rise to a (dynamical) background electric field. Eq.~\eqref{eq:hfceq} can be shown to solve \eqref{eq:beom} by inspection. On the other hand, the functions $U,\, V$ and $W$ can be determined by solving numerically 3 independent components of Einstein's equations:
\begin{subequations}
\label{eq:back_eom}
\begin{align}
    \label{eq:back_eom_start}
    & 2 V''+2 V'^2+W''+W'^2=0, \\
    & 2 \rho^2 e^{-4 V}+U' \left(V'-W'\right)+U \left(V''+\left(V'-W'\right) \left(2 V'+W'\right)-W''\right)=0, \\
    & 2 \rho^2 e^{-4 V}+2 U' V'+U' W'+4 U V' W'+2 U V'^2+\frac{1}{2} n ^2 e^{-4 V-2 W}-12=0.
    \label{eq:back_eom_end}
\end{align}
\end{subequations}
To find the resistivities $r_\parallel,~r_\perp$ and $r_H$,\footnote{\label{ft:nom}In this section, we always deal with electrical resistivities. Because of this, using this notation does not create any ambiguities.} it is sufficient to perturb the background with
\begin{align}
    & \delta B = \delta B_{ij}(r) e^{-i \omega t} dx^i\wedge dx^j, \nonumber \\[4pt]
    & \delta g = 2 e^{2V} \delta g_{ta}(r)e^{-i\omega t} dtdx^a + 2 e^{2W} \delta g_{tz}(r)e^{-i\omega t} dtdz,
\end{align}
where $a=x,y$. These can be decomposed into 2 channels, which in line with \Cref{sec:mhd} we call longitudinal and transverse (with respect to the background magnetic field, which points along $z$). The longitudinal channel is governed by:
\begin{subequations}
\label{eq:rpeom}
\begin{align}
    & n U \delta B_{{xy}}' - i \omega  e^{4V +2 W} \delta g_{tz}' = 0, \\
    & \delta B_{xy}'' + \left(\frac{U'}{U}-2 V'+ W'\right) \delta B_{xy}' +\frac{\omega ^2 \delta B_{xy}}{U^2} -\frac{i n  \omega  \delta g_{tz}}{U^2}  = 0.
\end{align}
\end{subequations}
The transverse one is more complicated:
\begin{subequations}
\label{eq:rteom}
\begin{align}
    & \delta B_{xz}'' + \left(\frac{U'}{U} - W'\right) \delta B_{xz}' + \frac{\omega^2}{U^2} \delta B_{xz} + \frac{i n \omega}{U^2} \delta g_{ty} + \frac{2 \rho e^W}{U} \delta g_{tx}' = 0, \\ 
    & \delta B_{yz}'' + \left(\frac{U'}{U} - W'\right) \delta B_{yz}' + \frac{\omega^2}{U^2} \delta B_{yz} - \frac{i n \omega}{U^2} \delta g_{tx} + \frac{2 \rho e^W}{U} \delta g_{ty}' = 0, \\
    & n U \delta B_{{yz}}' - i \omega  e^{4V +2 W} \delta g_{tx}' - 2 i \omega \rho e^W \delta B_{xz} + 2 n \rho e^W \delta g_{ty} = 0, \\
    & n U \delta B_{{xz}}' + i \omega  e^{4V +2 W} \delta g_{ty}' + 2 i \omega \rho e^W \delta B_{yz} + 2 n \rho e^W \delta g_{tx} = 0.
\end{align}
\end{subequations}
Notably, it is only in absence of $n$ that $\delta B_{xz}$ and $\delta B_{yz}$ decouple, leading to $r_H=0$. This marks a stark difference with the probe limit procedure in the previous \Cref{sec:grav_probe}, and is consistent with the fact that Hall resistivity must appear only if charge conjugation symmetry is broken, in this case, by the choice of the state. Another important observation, which we elaborate on in the following, is that in case $n\neq0$, there is an extra constant solution in addition to the one found numerically:
\begin{equation}
    \label{eq:rgdof}
    \omega \delta \bar{B}_{xy} = i n \delta \bar{g}_{tz}, \quad \omega \delta \bar{B}_{xz} = - i n \delta \bar{g}_{ty}, \quad \omega \delta\bar{B}_{yz} = i n \delta \bar{g}_{tx}.
\end{equation}
These constant modes have been first noted in \cite{Hartnoll:2007ai} and interpreted as a residual gauge degree of freedom in \cite{Donos:2013eha}. Their absence from the analysis of charged anisotropic plasmas presented in \cite{Demircik:2024bxd} is the reason why that work does not have a Hall sector.  

The remainder of this section consists of the systematic analysis of the numerical integration of each family of ODEs reported above. At the end, we calculate the transport coefficients and compare them with the probe limit results.

\subsection{Background}
\label{sec:num_bcg}

We start our analysis by numerically solving the background equations \eqref{eq:back_eom}. One peculiar aspect of the magnetic brane background is that the most convenient coordinates to solve the equations do not lead to an asymptotically AdS metric. For this reason, we will distinguish between \textit{numerical} and \textit{physical} coordinates when appropriate. As we elaborate later, their difference is in the overall scale of the boundary spatial coordinates $x,y$ and $z$. Numerical coordinates are defined by requiring:
\begin{equation}
    \label{eq:aachs}
    V(r_h) = W(r_h) = 0,
\end{equation}
whereas the existence of an event horizon implies that:
\begin{equation}
    U(r_h) = 0,
\end{equation}
for some radial coordinate $r_h$. Given that a unique solution can be specified by 5 parameters, there are two more that need to be fixed. By picking the scale of the radial coordinate, we can impose $r_h=1$. The last free parameter to be fixed is $U'(r_h)$. We fix this numerically by requiring the vanishing of a certain UV coefficient (see below).  

The actual integration is simplified by a couple of tricks. The first one is factorizing the divergent part of the unknown metric functions $U,\, V$ and $W$ by defining:
\begin{equation}
    U(r) = r^2 u(r), \quad V(r) = \ln r + v(r), \quad W(r) = \ln r +  w(r).
\end{equation}
The other is to change the radial coordinate to make the domain of integration compact. In this work, we choose:
\begin{equation}
    q = \frac{r_h}{r},
\end{equation}
where we recall that $r_h=1$. After these transformations, the (numerical) metric ansatz reads:
\begin{equation}
    \label{eq:qmet}
    ds^2 = \frac{1}{q^2} \spars{- u(q) dt^2 + \frac{dq^2}{u(q)} + e^{2 v(q)} (dx^2 + dy^2) + e^{2 w(q)} dz^2 }.
\end{equation}

The background functions will then be specified by integrating the coordinate transformed \eqref{eq:back_eom} from the horizon $q=1$ to the AdS boundary $q=0$. The UV ($q\to0$) expansion of the background functions, as determined from the equations of motion, reads:
\begin{align}
        & u(q) = 1 + u_{(1)} z + \frac{1}{4} u_{(1)}^2 q^2 + u_{(4)} q^4 + \frac{2}{3} \rho^2 e^{-4 v_{(0)}} q^4 \ln q + \mathcal{O}(q^5 \ln q), \nonumber\\
        & v(z) = v_{(0)} + \frac{1}{2} u_{(1)} q - \frac{1}{8} u_{(1)}^2 q^2 + \frac{1}{24} q^3 u_{(1)}^3 + v_{(4)} q^4 - \nonumber\\ & \hspace{.5\textwidth} - \frac{1}{6} \rho^2 e^{-4 v_{(0)}} q^4 \ln q + \mathcal{O}(q^5 \ln q), \nonumber\\
        & w(z) = w_{(0)} + \frac{1}{2} u_{(1)} q - \frac{1}{8} u_{(1)}^2 q^2 + \frac{1}{24} q^3 u_{(1)}^3 - \left(2 v_{(4)} + \frac{3}{64} u_{(1)}^4\right) q^4 + \nonumber\\ \label{eq:UVb_exp} &\hspace{.5\textwidth} + \frac{1}{3} \rho^2 e^{-4 v_{(0)}} q^4 \ln q + \mathcal{O}(q^5 \ln q).
\end{align}
The presence of the $u_{(1)}$ coefficient makes the UV expansion overly complicated, and difficult to compare with the Fefferman-Graham expansion needed to extract boundary data. Fortunately, this can be eliminated by taking an appropriate radial diffeomorphism \cite{Janiszewski:2015ura}. In practice, we implement it by requiring the condition
\begin{equation}
    \label{eq:num_cond}
    u_{(1)}(u'(1)) = 0,
\end{equation}
which is the last condition needed to fix the numerical coordinate system. After imposing the condition, the UV expansions of $u,\, v$ and $w$ simplify to
\begin{align}
    & u(q) = 1 +  u_{(4)} q^4 + \frac{2}{3} \rho_\text{phys}^2 q^4 \ln q + \frac{n_\text{phys}^2}{12} q^6 +  \mathcal{O}(q^7 \ln q), \nonumber \\
    & v(q) = v_{(0)} + v_{(4)} q^4 - \frac{1}{6} \rho_\text{phys}^2 q^4 \ln q + \mathcal{O}(q^7 \ln q), \nonumber \\
    \label{eq:UVb_expFG}
    & w(q) = w_{(0)} - 2 v_{(4)} q^4 + \frac{1}{3} \rho_\text{phys}^2 q^4 \ln q + \mathcal{O}(q^7 \ln q),
\end{align}
where we defined the physical charges
\begin{equation}
    \label{eq:phys_chg}
    \rho_\text{phys} = \rho e^{-2 v_{(0)}}, \quad n_\text{phys} = n e^{-2 v_{(0)}-w_{(0)}},
\end{equation}
which coincide with the boundary magnetic field and charge density, respectively.

\paragraph{Physical coordinate system}--- It is apparent from \eqref{eq:UVb_expFG} that because $(v(q),w(q)) \to (v_{(0)},w_{(0)})$ close to the UV boundary, the metric induced at the AdS boundary by \eqref{eq:qmet} is
\begin{equation}
    ds_4^2 = -dt^2 + e^{2v_{(0)}} (dx^2 + dy^2) + e^{2w_{(0)}} dz^2,
\end{equation}
the form of which (rather than the standard Minkowski metric) is a byproduct of the arbitrary choice of scale for $x,~y$ and $z$ implied by \eqref{eq:aachs}. This can be remedied by an appropriate additional rescaling, which defines the \textit{physical} coordinates:
\begin{equation}
    \label{eq:cntp}
    (x,y)_\text{phys} = e^{v_{(0)}}(x,y)_\text{num}, \quad z_\text{phys} = e^{w_{(0)}}z_\text{num}.
\end{equation}
In practice, this rescaling amounts to a translation of the functions $v$ and $w$ -- which sets their boundary value to 0 -- with the extra factors absorbed by the physical charges \eqref{eq:phys_chg}.

\paragraph{Probe limit transport coefficients}--- As a first application, we provide explicit expressions for the probe transport coefficients \eqref{eq:pb_tcs}. It is straightforward to prove, from the form of the metric \eqref{eq:qmet} and the relationship between the physical and numerical coordinates \eqref{eq:cntp}, that
\begin{equation}
    \label{eq:pblrs}
    r_\parallel = e^{2v_{(0)} - w_{(0)}}, \quad r_\perp = e^{w_{(0)}}, \quad r_H = 0,
\end{equation}
where we set $r_h=1$. In line with the findings of \Cref{sec:grav_probe}, the expression is valid both for $n=0$ and $n\neq0$. We note that due to the dependence of the UV coefficients $v_{(0)}$ and $w_{(0)}$ on $n$, this does not imply the probe resistivities to be equal in the two cases.

\paragraph{On the selection of charged\footnote{When referring to magnetic brane solutions, the words `neutral' and `charged' refer to $n$ rather than the magnetic charge $\rho$ of \Cref{sec:mhd}. While context should be enough to clarify any confusion, we will usually explicitly state which charge is being referred to.} (\boldmath$n\neq0$) backgrounds}--- The above procedure outlines how to select a unique background, given a choice of $\rho$ and $n$. As we clarified above, these are not however the boundary magnetic field and charge density. The latter have to be determined numerically by solving \eqref{eq:back_eom}, thus making the task of selecting the appropriate charged background difficult.

To see why, let us consider for a moment the neutral case. The only dimensionful parameters of the boundary theory are $T$ and $\rho_\text{phys}$. Because of scale invariance, any quantities will depend on the dimensionless ratio $t = T/\sqrt{\rho_\text{phys}}$ only. As we need to sample $t$ rather than picking a given value, the task here is simple.

In the charged case, we will need to keep another dimensionless ratio constant as we vary $t$. Our choice is:
\begin{equation}
    \label{eq:phad}
    \frac{n_\text{phys}}{\rho^{3/2}_\text{phys}} = \frac{n}{\rho^{3/2}} e^{v_{(0)}-w_{(0)}} = \const.
\end{equation}
Because of the dependence on $v_{(0)}$ and $w_{(0)}$, the ratio depends on boundary data -- which makes it necessary to fix either $n$ or $\rho$ via a root-finding method, each step of which would have to encapsulate the root-search \eqref{eq:num_cond}. In our analysis, we observed that the combination $v_{(0)}-w_{(0)}$ is independent of the choice of the initial condition of $u'(1)$, up to numerical precision. While we could not find a mathematical argument to justify it, this surprising finding allows to fix the ratio \eqref{eq:phad} before imposing the root condition \eqref{eq:num_cond} -- thus making the process much more streamlined and computationally efficient.

\subsubsection{Thermodynamic variables}
\label{sec:htd}

Before discussing transport coefficients, it is appropriate to elaborate on the analysis of thermodynamic quantities conducted in \cite{Grozdanov:2017kyl}. In \Cref{app:hrtq1} we derive the following formulae valid for magnetic brane solutions:
\begin{subequations}
    \begin{align}
        & \varepsilon = \thev{T_{tt}} = -\frac{3}{4}u_{(4)}+\frac{\rho^2_\text{phys}}{8\pi\bar\alpha}, \\
        & p = \thev{T_{aa}}\Big|_{a=x,y} = 2v_{(4)} -\frac{1}{4}u_{(4)}+\frac{\rho^2_\text{phys}}{4} \pars{\frac{1}{2\pi\bar\alpha}-1},
    \end{align}
\end{subequations}
which agree with the original reference once all differences in convention choices are taken into account. Here, $\bar\alpha$ is a free parameter arising from regularization of the logarithmic divergence of the on-shell matter sector of the action, corresponding to a the choice of characteristic scale. Importantly, the usual holographic renormalization procedure -- which prescribes the cancellation of all infinities and suppresses the dynamics of the boundary gauge fields -- corresponds to $\bar\alpha=\infty$, whereas the original reference \cite{Grozdanov:2017kyl} with dynamical electromagnetism (see also discussion in \cite{Grozdanov:2018fic}) chose $\bar\alpha=1/137$.

\begin{figure}
    \centering
    \includegraphics[width=\linewidth]{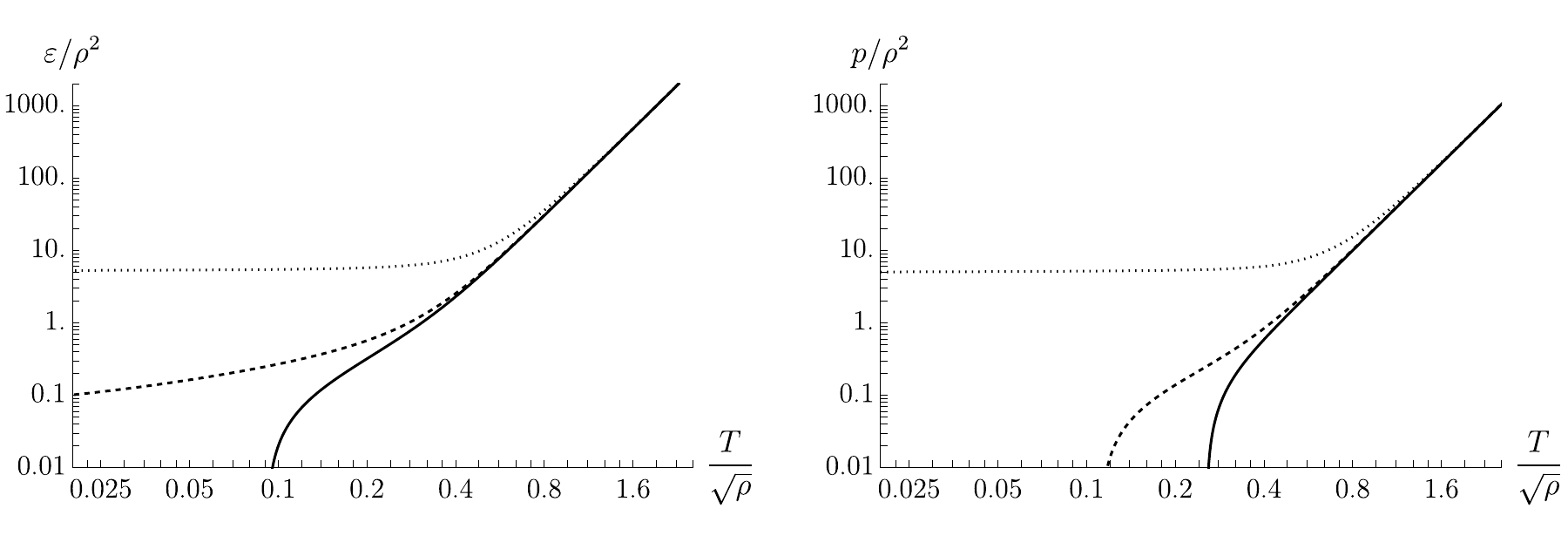}
    \caption{Energy density $\varepsilon$ and pressure $p$ for neutral magnetic branes ($n_\text{phys}=0$) as functions of dimensionless temperature $T/\sqrt{\rho_\text{phys}}$ for different values of the renormalization constant $\bar\alpha$: $\infty$ (solid), $1/2\pi$ (dashed) and $1/137$ (dotted). $\rho_\text{phys}$ is abbreviated to $\rho$ in the plots.}
    \label{fig:tdbcg_n}
\end{figure}

\begin{figure}
    \centering
    \includegraphics[width=\linewidth]{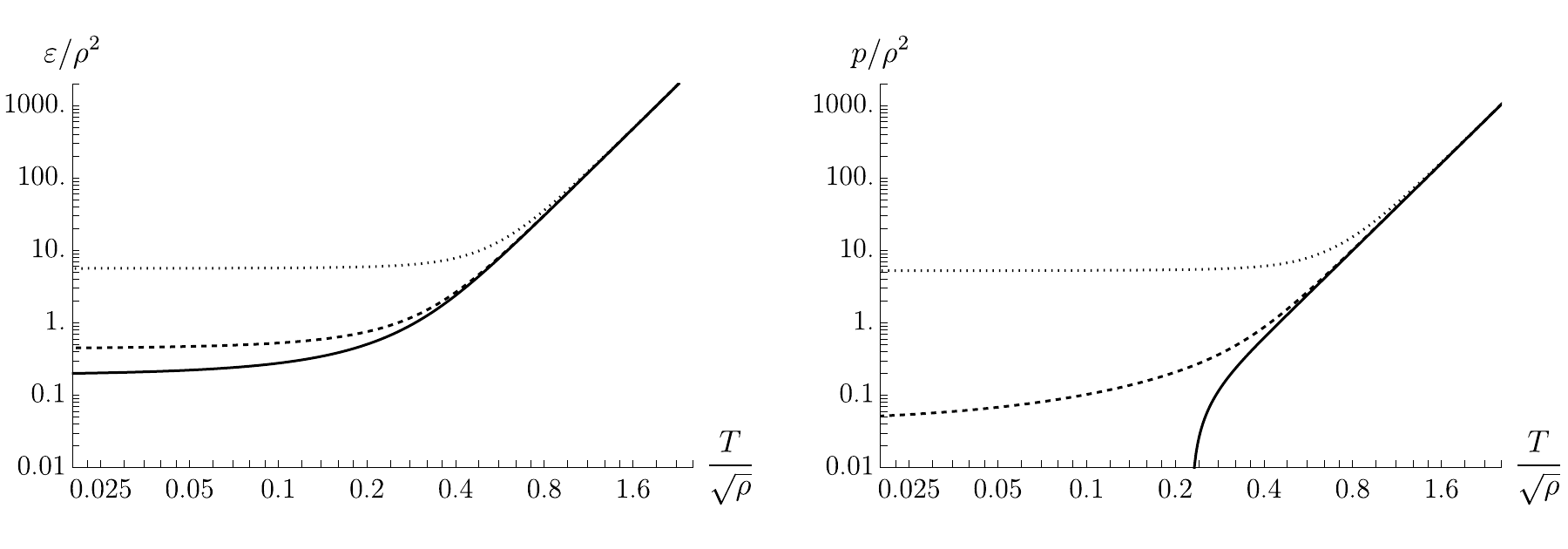}
    \caption{Energy density $\varepsilon$ and pressure $p$ for charged magnetic branes ($n_\text{phys}/\rho_\text{phys}^{3/2}=1$) as functions of dimensionless temperature $T/\sqrt{\rho_\text{phys}}$ for different values of the renormalization constant $\bar\alpha$: $\infty$ (solid), $1/2\pi$ (dashed) and $1/137$ (dotted). $\rho_\text{phys}$ is abbreviated to $\rho$ in the plots.}
    \label{fig:tdbcg_c}
\end{figure}

\begin{figure}
    \centering
    \includegraphics[width=\linewidth]{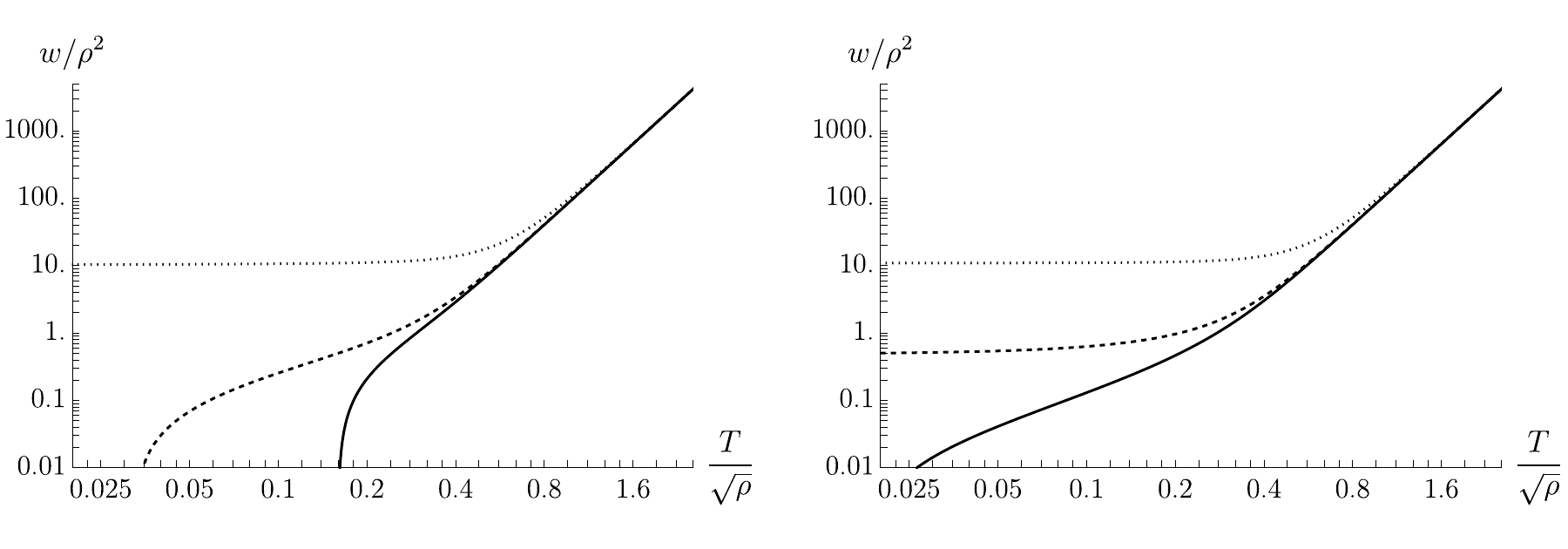}
    \caption{Enthalpy density $w=\varepsilon+p$ for neutral (left) and charged (right) magnetic branes  ($n_\text{phys}/\rho_\text{phys}^{3/2}=1$) as functions of dimensionless temperature $T/\sqrt{\rho_\text{phys}}$ for different values of the renormalization constant $\bar\alpha$: $\infty$ (solid), $1/2\pi$ (dashed) and $1/137$ (dotted). $\rho_\text{phys}$ is abbreviated to $\rho$ in the plots.}
    \label{fig:enths_comp}
\end{figure}

To better understand the impact of the choice of renormalization scale on these quantities, we study their dependence on dimensionless temperature for different values of $\bar\alpha$, both at vanishing and at non-zero charge density $n_\text{phys}$.
The results are respectively reported in \Cref{fig:tdbcg_n,fig:tdbcg_c}. In the $n_\text{phys}=0$ case we observe that when the choice of $\bar\alpha$ is the same, our results match those in \cite{Grozdanov:2017kyl} up to numerical precision. Perhaps more interestingly, larger values of $\bar\alpha$ lead to a more pronounced decrease of $\varepsilon$ and $p$ with temperature, making them negative in some cases. A similar picture emerges in the $n_\text{phys}\neq0$, though our analysis suggests that the presence of a finite charge density has an overall stabilizing effect.

The simultaneous decrease in energy and pressure may eventually lead to a zero-enthalpy point $w=\varepsilon+p=0$, which is indicative of a breakdown of the hydrodynamic approximation and a signature of a phase transition. To investigate this possibility, we plot the enthalpy density in the two cases under analysis in \Cref{fig:enths_comp}. We find evidence of a $w=0$ point for larger values of $\bar\alpha$ in the $n_\text{phys}=0$ case, though our numerical data does not exclude that at very low temperatures even smaller values could develop such a feature. The analysis of the $n_\text{phys}\neq0$ case is less conclusive, as we cannot ascertain the existence of a $w=0$ even at $\bar\alpha=\infty$.

\paragraph{A magnetic BF bound}--- In the case of $n_\text{phys}=0$, the question on the existence of the $w=0$ point for arbitrary values of $\bar\alpha$ can be addressed by deriving an upper bound on the magnetic field in the extremal solution. This line of reasoning mirrors the derivation of the Breitenlohner-Freedman bound presented in \cite{Moroz:2009kv}, and is somewhat reminiscent of the investigation of the superconducting transition at zero gauge coupling in \cite{Hartnoll:2008kx}. For this reason, we refer to this bound as a `magnetic BF bound'.

We start by observing that at zero temperature and  $n_\text{phys}=0$, $U=e^{2W}$ in \eqref{eq:metr} \cite{DHoker:2009mmn}. With some algebra we can derive an equation for $\delta B_{xz}$:
\begin{equation}
    \delta B_{xz}'' + \frac{U'}{2U} \delta B_{xz} + \pars{\frac{\omega^2}{U^2} - \frac{4 \rho^2 e^{4V}}{U} }\delta B_{xz} = 0.
\end{equation}
By changing radial coordinates to $q$ and using the UV expansions \eqref{eq:UVb_expFG}, we find that close to the boundary the equation above becomes:
\begin{equation}
    \delta B_{xz}'' + \frac{1}{q} \delta B_{xz}' + \pars{\omega^2-\frac{4 \rho^2_\text{phys}}{q^2}} \delta B_{xz} = 0.
\end{equation}
This equation can be put in Schrödinger form via the substitution: $\delta B_{xz} \to q^{-1/2}\delta b$. After this, we find
\begin{equation}
    -\delta b'' + \Phi(q) \delta b = \omega^2 \delta b, \quad \Phi(q) = - \frac{1}{q^2} (1 - 4\rho_\text{phys}^2).
\end{equation}
The inverse square potential has a ground state only under the condition \cite{Essin:2006sic}:
\begin{equation}
    \label{eq:magb}
    1 -4\rho_\text{phys}^2 < \frac{1}{4} \quad\Rightarrow\quad \rho_\text{phys}<\frac{\sqrt3}{4}.
\end{equation}
The absence of a ground state maps to a bulk gravitational instability \cite{Moroz:2009kv}. By inspection, it is easy to see that at very low temperatures, the bound \eqref{eq:magb} is violated. Thus, we interpret the appearance of a zero enthalpy point as follows: at low enough temperatures, the magnetic field causes an instability, which leads to a phase transition. This phase transition is captured on the boundary by the zero enthalpy point: hence, we conclude that for any value of $\bar\alpha$ one would expect to find such a point. It just so happens that for low enough values of $\bar\alpha$, $T(w=0)$ becomes increasingly difficult to observe by using numerical methods.

\subsection{Longitudinal channel}

The longitudinal channel is the simplest in that only 2 coupled equations need to be solved. In $q$-coordinates, the EOMs \eqref{eq:rteom} are:
\begin{subequations}
\label{eq:eqpc0}
\begin{align}
    & n u q^4 \delta B_{{xy}}' - i \omega  e^{4v +2 w} \delta g_{tz}' = 0, \\
    \label{eq:epqc}
    & \delta B_{xy}'' + \left(\frac{1}{q}+\frac{u'}{u}-2 v'+ w'\right) \delta B_{xy}' +\frac{\omega ^2 \delta B_{xy}}{u^2} -\frac{i n \omega  \delta g_{tz}}{u^2}  = 0.
\end{align}
\end{subequations}
The holographic Kubo formulae for the model \eqref{eq:uberS} have been given in \cite{Frangi:2024enh}:
\begin{equation}
    \label{eq:hk_par}
    r_\parallel = \lim_{\omega\to0}\frac{\delta B_{xy}^{(1)}}{i \omega (\delta B_{xy}^{(0)} -  \frac{1}{4 \pi \bar\alpha} \delta B_{xy}^{(1)})} \Bigg|_{\delta g_{tz}^{(0)}=0},
\end{equation}
where the coefficients are the ones appearing in the physical $q$-coordinate UV expansion:
\begin{align}
    & \delta B_{xy}(q) = \delta B_{xy}^{(0)} + \delta B_{xy}^{(1)} \ln q + \mathcal{O}(q^2 \ln q), \nonumber \\
    \label{eq:apcfs}
    & \delta g_{tz}(q) = \delta g_{tz}^{(0)} + \frac{i n_\text{phys}}{4\omega} \delta B_{xy}^{(1)} q^4 + \mathcal{O}(q^6).
\end{align}
All that is left to do is to numerically integrate \eqref{eq:eqpc0} from the horizon to the UV boundary, imposing ingoing boundary conditions at the former, and to read off the coefficients in \eqref{eq:apcfs} from the asymptotic behavior of the solutions. Because the details regarding this procedure significantly overlap with the treatment presented in \cite{Frangi:2024enh}, we refer the interested reader to that work for further detail.

\paragraph{Neutral (\boldmath$n=0$) magnetic branes}--- For neutral magnetic branes $n=0$, $\delta g_{tz}$ and $\delta B_{xy}$ decouple, and it is possible to derive an analytic result for $r_\parallel$. Because of the DC limit in \eqref{eq:hk_par}, we only need to know $\delta B_{xy}^{(1)}$ to linear order in $\omega$. Solving the neutral \eqref{eq:epqc} at that order one finds:
\begin{equation}
    \delta B_{xy}(q) = \tilde c \pars{1 + c \int_1^q dp \frac{e^{2 v(p) - w(p)}}{p u(p)}},
\end{equation}
where $c, \tilde c$ are constants fixed by imposing boundary conditions. At the event horizon, these read
\begin{equation}
    \delta B_{xy}(q) \xrightarrow{q\to1} \tilde c \, (1 - q)^{i \omega / 4 \pi T} \approx \tilde c \spars{ 1 + \frac{i\omega}{4\pi T} \ln(1-q) }.
\end{equation}
It follows that, to linear order in $\omega$,
\begin{equation}
    c = i \omega e^{w(1) - 2v(1)}.
\end{equation}
The value of $v$ and $w$ at $q=1$ is effectively fixed by the rescaling of the spatial coordinates \eqref{eq:cntp}, which leads to
\begin{equation}
    v(1) = -v_{(0)}, \quad w(1) = -w_{(0)},
\end{equation}
that can then be inserted into \eqref{eq:hk_par} to give
\begin{equation}
    r_\parallel = e^{2v_{(0)} - w_{(0)}}.
\end{equation}
This matches the known result of \cite{Grozdanov:2017kyl} and the probe limit expression \eqref{eq:pblrs}. Rather surprisingly, because $\delta B_{xy}^{(1)} = \mathcal{O}(\omega)$, we expect no dependence on $\bar\alpha$.

\paragraph{Charged (\boldmath$n\neq0$) magnetic branes}--- For charged magnetic branes, no simple analytic result is available, and one  needs to compute the Kubo formula \eqref{eq:hk_par}. One important aspect to notice is that the equations of motion are invariant under the following internal-space translation:
\begin{equation}
    \label{eq:dbs1}
    \delta B_{xy} \to \delta B_{xy} + c, \quad \delta g_{tz} \to \delta g_{tz} - \frac{i \omega}{n} c.
\end{equation}
This symmetry is due to the presence of residual gauge freedom, corresponding to the constant solutions \eqref{eq:rgdof} and is generated by the small diffeomorphism $\xi^\ind{A} = \delta^\ind{A}_q \xi e^{-i \omega t}$, with $\xi$ a constant. As the Kubo formula \eqref{eq:hk_par} clarifies, one needs to impose the condition $\delta g_{tx}^{(0)}=0$ before computing $r_\parallel$. Rather than imposing it numerically with a root-finding method, we can simply take advantage of \eqref{eq:dbs1} and subtract it at the end. We find
\begin{equation}
    \label{eq:hk_parz}
    r_\parallel = \lim_{\omega\to0}\frac{\delta B_{xy}^{(1)}}{i \omega \left(\delta B_{xy}^{(0)} -  \frac{1}{4 \pi \bar\alpha} \delta B_{xy}^{(1)} - \frac{i n_\text{phys}}{\omega} \delta g_{tz}^{(0)}\right)},
\end{equation}
without extra constraints. In our numerical computations we adopt this efficient strategy.

\begin{figure}
    \centering
    \includegraphics[width=\linewidth]{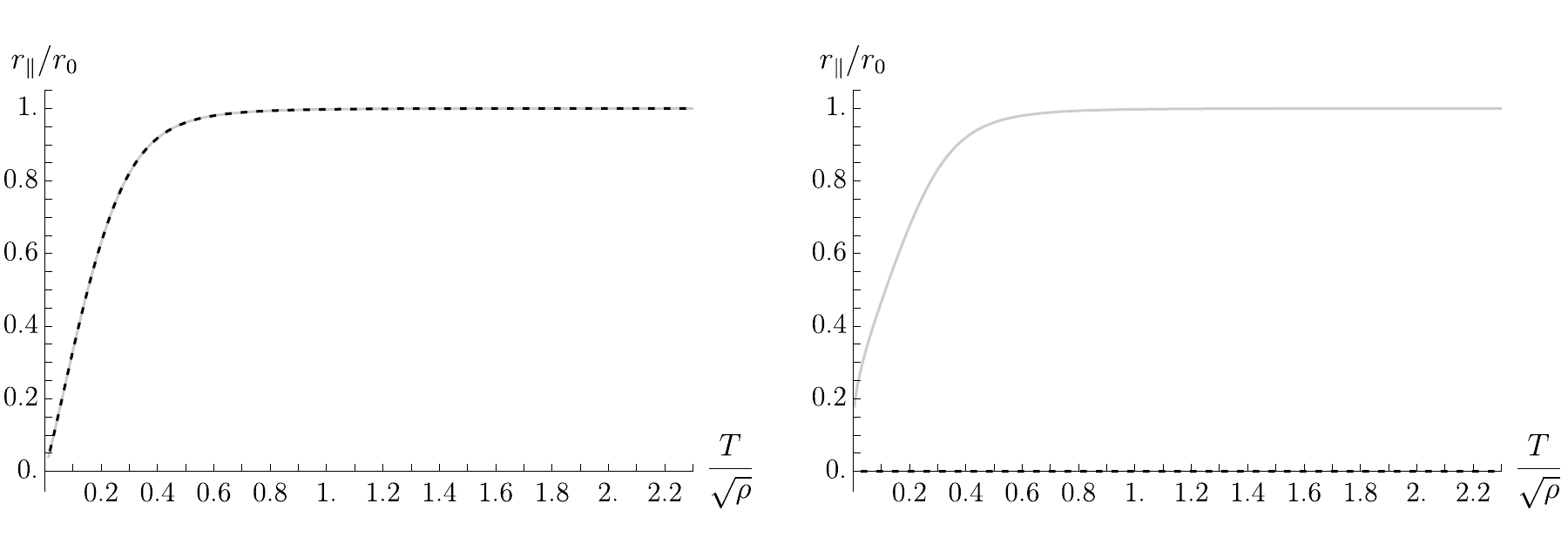}
    \caption{Parallel resistivity for neutral (left) and charged (right, $n_\text{phys}/\rho^{3/2}_\text{phys}=1$) magnetic branes, as a function of dimensionless temperature. The transport coefficients are normalized using the resistivity for a Schwarzschild black brane $r_0$. The gray lines correspond to the probe limit expressions \eqref{eq:pblrs}, whereas the dashed black lines to the values calculated via \eqref{eq:hk_parz}.}
    \label{fig:parallel}
\end{figure}

\paragraph{Results and comments}--- The above provides the ingredients for a numerical calculation of the parallel resistivity $r_\parallel$ for both neutral and charged magnetic branes via \eqref{eq:hk_parz}, to be compared with the probe limit expression in \eqref{eq:pblrs}. We sum this up in \Cref{fig:parallel}.

As anticipated, for $n=0$ we find perfect agreement between the probe limit formula -- which is entirely expressed in terms of background quantities -- and the values extracted by studying the linearized perturbations. This is not the case when $n\neq0$. In particular, the probe limit formula misses the invariance of the solution under boundary spatial translations. The vanishing of the resistivity as a consequence of this property, which forces every electric current to be a `supercurrent', has been reported in \cite{Frangi:2024enh}, and follows established facts about holographic conductivities (see \cite{Hartnoll:2016apf} for a comprehensive review of the argument). To conclude, we find that the value of $\bar\alpha$ does not have any effect on the transport coefficients, confirming our previous findings.

\subsection{Transverse channel}
\label{sec:transv}

While the longitudinal sector outlined some of the shortcomings of the gravitational probe limit, it is in the transverse channel that the distinction between probe limits becomes most prominent. Once again, we start from the $n=0$ case -- corresponding to the hydrodynamic theory studied in \Cref{sec:mhd}, and consider the more complicated $n\neq0$ setting later.

\paragraph{Neutral magnetic branes}--- When $n=0$, the equations of motion describing the dynamics of the transverse sector decouple in pairs. In $q$ coordinates one finds:
\begin{subequations}
    \begin{align}
        & \delta B_{xz}'' + \pars{\frac{1}{q} + \frac{u'}{u} - w'} \delta B_{xz}' + \frac{\omega^2}{u^2} \delta B_{xz} - \frac{2 \rho e^w}{q u} \delta g_{tx}' = 0, \\[4pt]
        \label{eq:trn2eq}
        & \delta g_{tx}' - 2 \rho e^{-4v-w} q^3 \delta B_{xz} =0,
    \end{align}
\end{subequations}
and an analogous couple obtained by substituting: $(\delta B_{xz},\delta g_{tx}) \to (\delta B_{yz},\delta g_{ty})$. The decoupling between $\delta B_{xz}$ and $\delta B_{yz}$ implies that $r_H=0$, in accordance with \cite{Vardhan:2022wxz}. Incidentally, the second equation allows to eliminate any reference to metric fluctuations from the first, leading to:
\begin{equation}
    \label{eq:dtrn}
    \delta B_{xz}'' + \pars{\frac{1}{q} + \frac{u'}{u} - w'} \delta B_{xz}' + \pars{ \frac{\omega^2}{u^2} - \frac{4 \rho^2 q^2 e^{-4v}}{u} } \delta B_{xz}  = 0.
\end{equation}
We should not confuse this decoupling with a probe limit, since \eqref{eq:trn2eq} clearly shows that for any nontrivial $\delta B_{xz}$ profile, $\delta g_{tx}\neq0$. For the sake of concreteness, the probe limit in the sense of \Cref{sec:grav_probe} corresponds to setting $\delta g_{\mn}=0$ and ignoring \eqref{eq:trn2eq} altogether. This would lead to:
\begin{equation}
    \label{eq:dtrne}
    \delta B_{xz}'' + \pars{\frac{1}{q} + \frac{u'}{u} - w'} \delta B_{xz}' + \frac{\omega^2}{u^2} \delta B_{xz}  = 0,
\end{equation}
from which it is possible, following the logic outlined above for the longitudinal sector, to analytically find
\begin{equation}
    r_{\perp} = e^{w_{(0)}},
\end{equation}
in accordance with the expectation set by the probe limit effective action. 

In the transverse plane, the Kubo formula receives a correction from the nonzero fluid velocity in the direction perpendicular to electric transport \eqref{eq:ri3kl}. Accordingly, the holographic Kubo formula receives a correction:
\begin{equation}
    \label{eq:hk_trn}
    r_\perp = \lim_{\omega\to0} \spars{ \frac{\delta B_{xz}^{(1)} }{i \omega \pars{ \delta B_{xz}^{(0)} - \frac{1}{4 \pi \bar\alpha} \delta B_{xz}^{(1)} }} + \frac{\rho_\text{phys}^2}{i \omega(\varepsilon + p)} }_{\delta g_{ty}^{(0)}=0}.
\end{equation}
The form of this correction can be found on purely holographic grounds (via the asymptotic expansion of $\delta g_{ty}$), once a correction coming from the fluid velocity is allowed. Because all the background quantities entering the correction are real, we observe that all it amounts to is to cancel the imaginary part of the term on the left. Considering that an exact cancellation would require infinite numerical precision (which we cannot access), one practical way to check this is to consider their ratio instead of their subtraction. We find that for all values of $\bar\alpha$ considered, deviations of this ratio from 1 are $\mathcal{O}(10^{-6})$ or lower. In practice, we then consider the real part of \eqref{eq:hk_trn} when calculating the transverse resistivity. Additionally, it is worth noting that because $\delta g_{ty}$ enters the equations of motion \eqref{eq:trn2eq} through its derivatives, the vanishing source condition is completely trivial to implement.

\begin{figure}
    \centering
    \includegraphics[width=\linewidth]{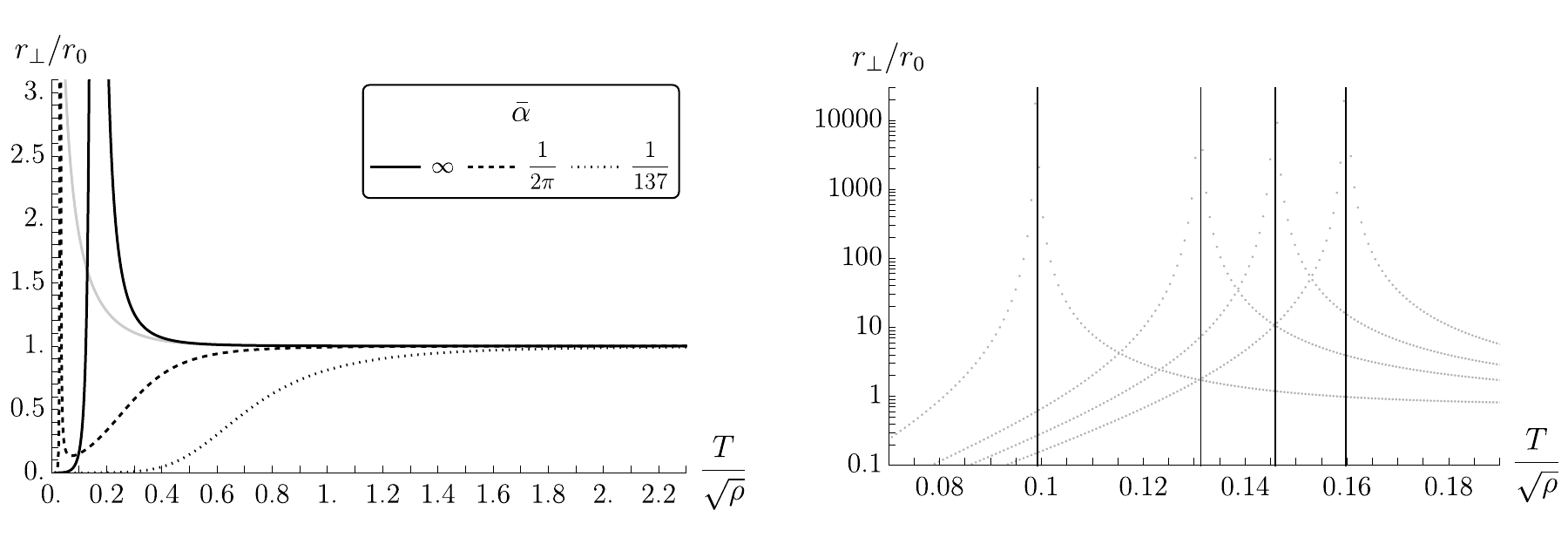}
    \caption{\textbf{Left}: Transverse resistivity for neutral magnetic branes, as function of dimensionless temperature, for various values of $\bar\alpha$, compared to the appropriate probe limit expression (gray). \textbf{Right:} Log-scale representation of the divergence of $r_\perp$ (gray points) and zero-enthalpy temperature (black lines) for $\bar\alpha \in \{\infty,1/\pi,1/2\pi,1/4\pi \}$ from right to left. The transport coefficients are normalized using the resistivity for a Schwarzschild black brane $r_0$.}
    \label{fig:trn_resn}
\end{figure}
We present our results in \Cref{fig:trn_resn}. We observe that for large values of $\bar\alpha$ the resistivity diverges at a finite temperature, in correspondence with the zero enthalpy point. This feature is compatible with the interpretation we gave in \Cref{sec:htd}, and reinforces the idea that at low enough temperatures, a phase transition is triggered by breaking the magnetic BF bound derived above. We stress that the hydrodynamic interpretation of the theory is still valid away from this point, and observe that for smaller values of $\bar\alpha$, no instability is observed -- though the existence of the bound suggests that it may nevertheless be present at sufficiently low temperatures.

\paragraph{Charged magnetic branes}--- In the $n\neq0$ case, no decoupling occurs, and one needs to solve the 4 coupled equations together. In $q$ coordinates these are:
\begin{subequations}
\label{eq:ctreqq}
    \begin{align}
        & \delta B_{xz}'' + \pars{\frac{1}{q} + \frac{u'}{u} - w'}\delta B_{xz}' + \frac{\omega^2}{u^2} \delta B_{xz} - \frac{2 \rho e^w}{q u} \delta g_{tx}' + \frac{i n \omega}{u^2} \delta g_{ty} = 0, \\[4pt]
        & \delta B_{yz}'' + \pars{\frac{1}{q} + \frac{u'}{u} - w'}\delta B_{yz}' + \frac{\omega^2}{u^2} \delta B_{yz} - \frac{2 \rho e^w}{q u} \delta g_{ty}' - \frac{i n \omega}{u^2} \delta g_{tx} = 0, \\[4pt]
        & 2 i \rho \omega q^3 \delta B_{xz} - 2 \rho n q^3 \delta g_{ty} + n q^4 u e^{-w} \delta B_{yz}' - i \omega e^{4 v + w} \delta g_{tx}' =0, \\[4pt] 
        & 2 i \rho \omega q^3 \delta B_{yz} + 2 \rho n q^3 \delta g_{tx} - n q^4 u e^{-w} \delta B_{xz}' - i \omega e^{4 v + w} \delta g_{tx}' = 0.
    \end{align}
\end{subequations}
The details of the numerical analysis do not significantly differ from what previously discussed. The Kubo formulae for $r_\perp$ and $r_H$ (which we may expect on the grounds of the coupling between $\delta B_{xz}$ and $\delta B_{yz}$) read:
\begin{subequations}
\label{eq:kctr}
    \begin{align}
        \label{eq:kctrp}
        & r_\perp = \lim_{\omega\to0} \left[ \frac{\delta B_{xz}^{(1)} }{i \omega \pars{ \delta B_{xz}^{(0)} - \frac{1}{4 \pi \bar\alpha} \delta B_{xz}^{(1)} + \frac{i n_\text{phys}}{\omega} \delta g^{(0)}_{ty} }} + \frac{\rho_\text{phys}^2}{i \omega(\varepsilon + p)} \right]_{E_x(\delta B_{yz}^{(0)},\delta B_{yz}^{(1)},\delta g^{(0)}_{tx})=0} \\
        \label{eq:kctrh}
        & r_H = \lim_{\omega\to0} \spars{\frac{- \delta B_{yz}^{(1)} }{i \omega \pars{ \delta B_{xz}^{(0)} - \frac{1}{4 \pi \bar\alpha} \delta B_{xz}^{(1)} + \frac{i n_\text{phys}}{\omega} \delta g^{(0)}_{ty} }} }_{E_x(\delta B_{yz}^{(0)},\delta B_{yz}^{(1)},\delta g^{(0)}_{tx})=0}
    \end{align}
\end{subequations}
where $E_x$ is the following linear combination:
\begin{equation}
    E_x(\delta B_{yz}^{(0)},\delta B_{yz}^{(1)},\delta g^{(0)}_{tx}) = \delta B_{yz}^{(0)} - \frac{1}{4 \pi \bar\alpha} \delta B_{yz}^{(1)} - \frac{i n_\text{phys}}{\omega} \delta g^{(0)}_{tx}.
\end{equation}
Defining the transport coefficients in this way fixes the gauge redundancy of the modes~\eqref{eq:rgdof}, analogously to what we did in \eqref{eq:hk_parz}. By imposing ingoing boundary conditions at the horizon, one finds that numerical solutions of \eqref{eq:ctreqq} are uniquely defined by two parameters. One can be fixed to unity due to scaling symmetry, whereas the other is fixed by requiring $E_x=0$ on the boundary. 

We observe that the physical interpretation of $E_x$ is that of an external electric field on the $x$ direction (hence the choice of notation). This suggests a way to avoid enforcing the condition via a root-finding method: rather than forcing the external source to be aligned with the $y$ direction, we could simply calculate the response in the parallel and perpendicular directions to the vector $(E_x,E_y)$, thus finding, respectively, $r_\perp$ and $r_H$. We follow this strategy.

\begin{figure}
    \centering
    \includegraphics[width=\linewidth]{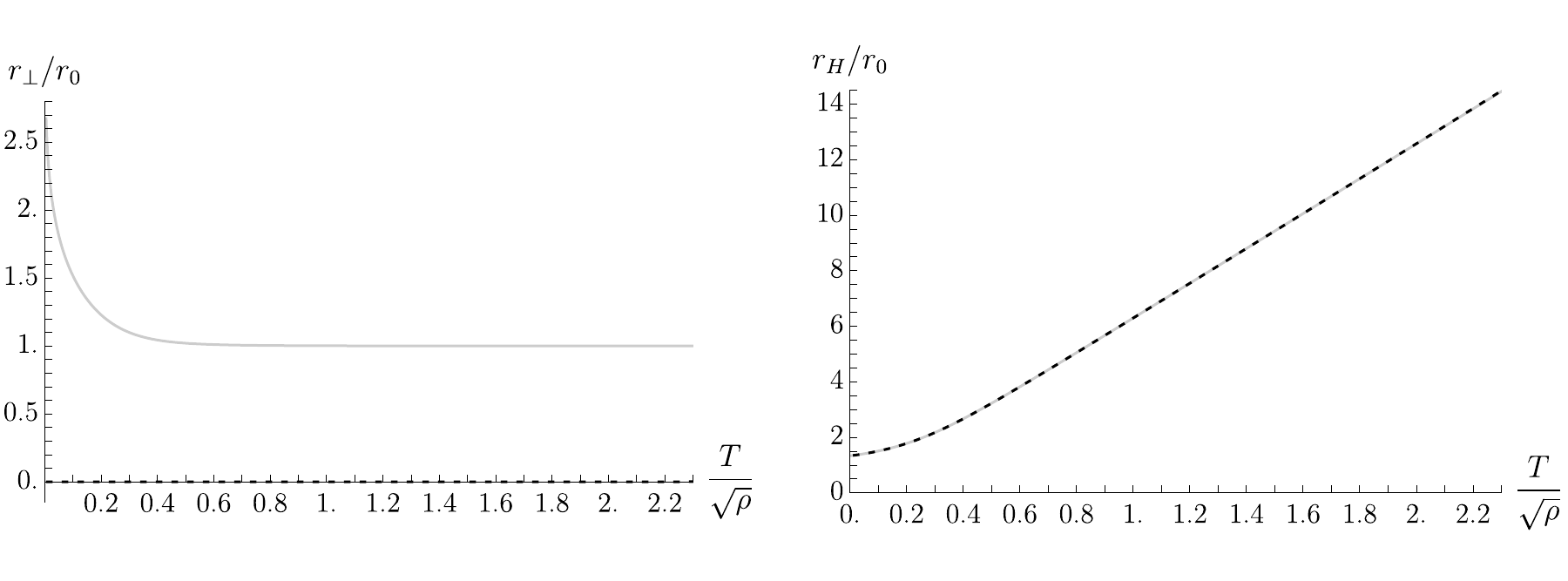}
    \caption{\textbf{Left}: Transverse resistivity (black, dashed) for charged magnetic branes ($n_\text{phys}/\rho_\text{phys}^{3/2}=1$), as function of dimensionless temperature, compared to the appropriate probe limit expression (gray). \textbf{Right:} Hall resistivity (black, dashed) for charged magnetic branes ($n_\text{phys}/\rho_\text{phys}^{3/2}=1$), as function of dimensionless temperature, compared to the analytic value \eqref{eq:rhan} (gray). The transport coefficients are normalized using the resistivity for a Schwarzschild black brane $r_0$.}
    \label{fig:transverse_c}
\end{figure}
We report the results of the numerical calculations in \Cref{fig:transverse_c}. Interestingly, there appears to be no dependence on $\bar\alpha$ in either $r_\perp$ or $r_H$. While this may seem odd at first, it is expected on symmetry grounds. That $r_\perp=0$ follows the intuition that in presence of a finite density of electric charge, all diagonal components of the resistivity must vanish. Additionally, in any Lorentz-invariant theory with a finite magnetic field (line density) $\rho_\text{phys}$ and charge density $n_\text{phys}$ the Hall resistivity is fixed to the value \cite{Hartnoll:2007ai} (see also \cite{Hernandez:2017mch} for a discussion of this fact in charged MHD):
\begin{equation}
    \label{eq:rhan}
    r_H = \frac{\rho_\text{phys}}{n_\text{phys}} = \frac{\rho}{n}e^{w_{(0)}}. 
\end{equation}
We see that, up to rounding errors, the prediction is confirmed by our numerical calculation.

\section{Implications for magnetic transport}
\label{sec:neutron}

In this final section, we discuss the phenomenological differences between the probe limit and full MHD. In particular, we focus on different dispersion relations in each regime. We derived their explicit expressions from hydrodynamic EFT arguments in \Cref{sec:mhd}, and holography provides us with explicit microscopically computed transport coefficients and thermodynamic quantities (\Cref{sec:grav_probe,sec:beyond}). Importantly, the dispersion relations depend on the magnetic viscosities, and we will have to be careful about how they differ from the electrical resistivities we calculated in \Cref{sec:beyond}. Throughout this section we set the electric charge to zero, $n_\text{phys}=0$, focusing on the neutral magnetic brane solution corresponding to the hydrodynamic theory of \Cref{sec:mhd}. We also drop the subscript `phys' in the remainder of this discussion and set $\bar\alpha=1/137$.

\begin{figure}
    \centering
    \includegraphics[width=.5\linewidth]{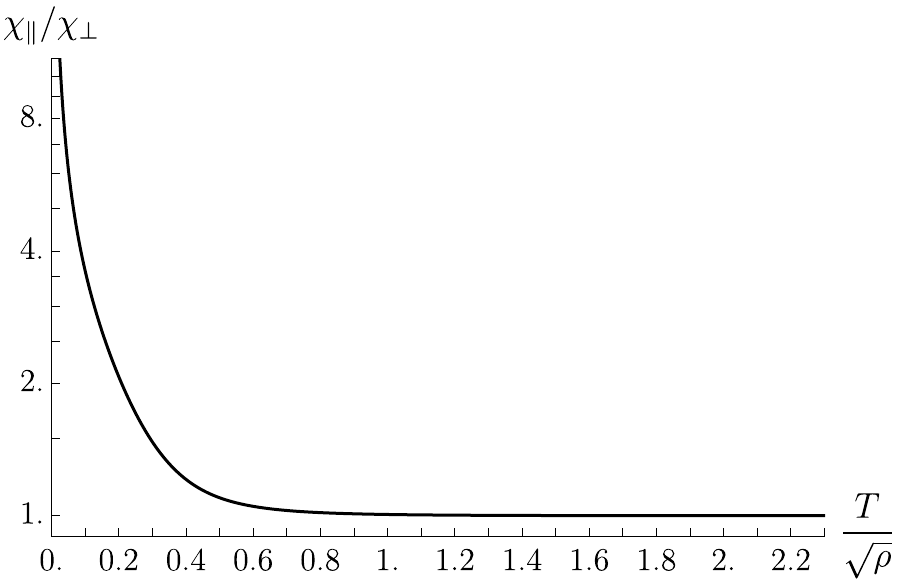}
    \caption{Ratio of the susceptibilities $\chi_\parallel$ and $\chi_\perp$ as a function of dimensionless temperature $t=T/\sqrt\rho$. }
    \label{fig:chi_comp}
\end{figure}

In order to characterize the modes, we need expressions for the entropy density, $s$, chemical potential, $\mu$, and shear viscosities, $\eta_\parallel$ and $\eta_\perp$, which were outside of the scope of the previous sections. These can be recovered from \cite{Grozdanov:2017kyl} once differences in conventions and normalizations are resolved:\footnote{In particular: $e^{v_{(0)}}\leftrightarrow v$, $e^{w_{(0)}}\leftrightarrow w$, and each factor of $N_c^2/2\pi^2$ appearing there should be set to 1.}
\begin{equation}
    s = \pi e^{-2 v_{(0)} - w_{(0)}}, \quad \mu = 6 v_{(4)} + \frac{\rho^2}{4}\pars{\frac{1}{\pi\bar\alpha} - 1}, \quad \eta_\parallel = \frac{s}{4\pi} e^{2 (v_{(0)}-w_{(0)})}, \quad \eta_\perp = \frac{s}{4\pi}.
\end{equation}
We introduce the following notation for the magnetic susceptibilities:
\begin{equation}
    \chi_\parallel = \pars{\frac{\partial\rho}{\partial\mu}}_T ,\quad \chi_\perp = \frac{\rho}{\mu}.
\end{equation}
In the context of most astrophysical neutron star calculations (see \cite{Vardhan:2022wxz} and \cite{goldreich1992magnetic}), the two susceptibilities are taken to be equal, $\chi_\parallel = \chi_\perp $, which is in part due to the fact that the magnetic field dependence of the equation of state of nuclear matter is not well understood. On the other hand, in holographic toy models, understanding the differences between the two susceptibilities is straightforward. The only subtlety to be taken care of is that $\mu$ is calculated numerically as a function of the dimensionless ratio $t = T/\sqrt{\rho}$, due to scale invariance on the boundary. Assuming the temperature $T$ to be constant, it is a simple change of variable that gives:
\begin{equation}
    \chi_\parallel = \pars{\frac{\partial \mu}{\partial \rho}}_T^{-1} = - \frac{2 \rho}{t \partial_t \mu(t)}.
\end{equation}
Given this expression, one can quantify how substantial the distinction between the two susceptibilities is (see also \cite{Vardhan:2022wxz}). We plot their ratio in \Cref{fig:chi_comp}, showing that while for weak magnetic fields ($t\gtrsim 1$) the difference is almost negligible, it becomes extremely important at strong magnetic fields ($t\lesssim1$). For this reason, we will consider these two cases separately for the remainder of this section.

\paragraph{Probe limit}--- In the probe limit and with spatial momentum parameterized as $\mathbf k = \kappa (\sin\theta,0,\cos\theta),$ there are two diffusive modes, given in \eqref{eq:pbmd2} by
\begin{equation}
    \label{eq:pb_disp}
    \omega_1 = - \frac{i \kappa^2}{\chi_\perp} (r_\perp \cos^2\theta + r_\parallel \sin^2\theta ), \quad
    \omega_2 = - i r_\perp \kappa^2 \pars{\frac{ \cos^2\theta}{\chi_\perp} + \frac{\sin^2\theta}{\chi_\parallel}}.
\end{equation}
\begin{figure}
    \centering
    \includegraphics[width=\linewidth]{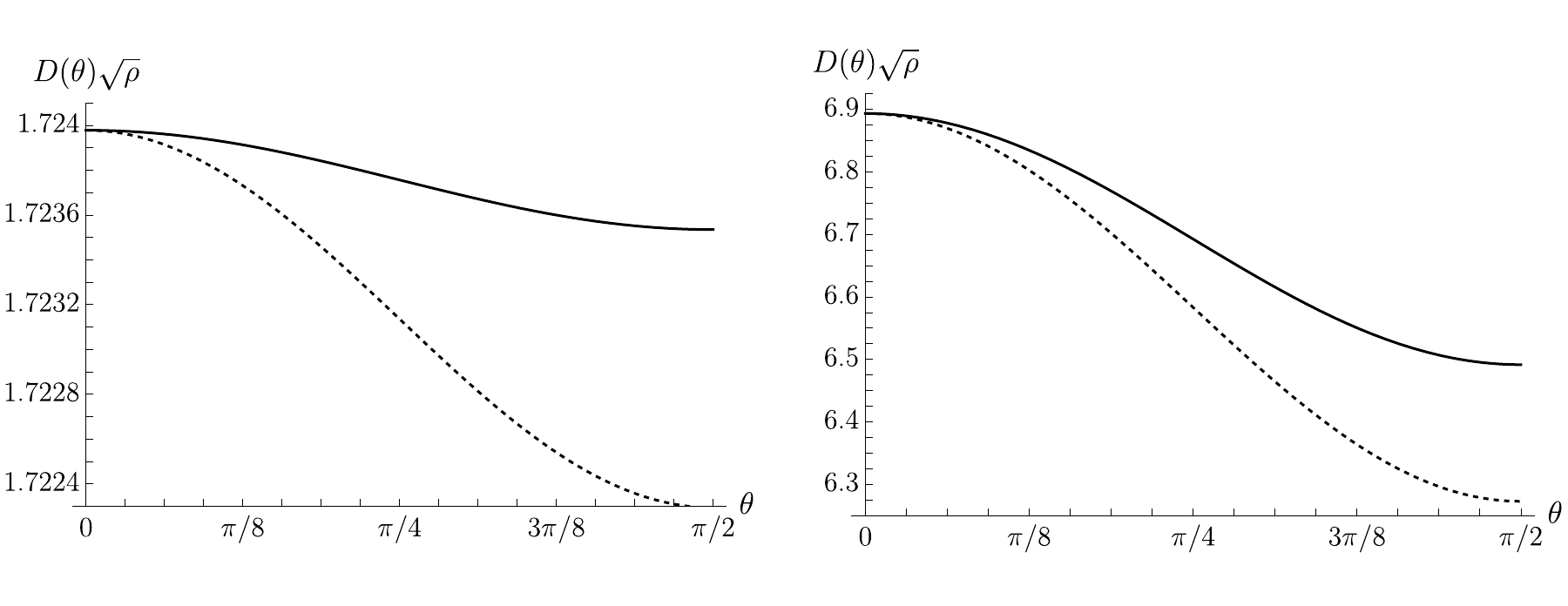}
    \caption{Diffusivities in the probe limit \eqref{eq:pb_disp}, with transport coefficients calculated from the holographic probe limit \eqref{eq:pblrs}, in the weak (left, $t\simeq 2$) and strong (right, $t\simeq0.5$) magnetic field regimes. $D_1(\theta)$ is the solid line, $D_2(\theta)$ the dashed one.}
    \label{fig:disp_probe}
\end{figure}

We define `diffusivities' $D_i(\theta) = i \omega_i / \kappa^2$ and plot them as functions of $\theta$ for two values of $t = T/\sqrt{\rho}$  in \Cref{fig:disp_probe}, so to visualize how the weak and strong magnetic field regimes behave differently. We start by noticing that due to the vanishing of Hall resistivity, these diffusivities are real in the whole $\theta\in[0,\pi/2]$ interval. Another observation is that even in the weak field case, $D_2$ is not constant, as it would be expected in the case $\chi_\parallel = \chi_\perp$.

\begin{figure}
    \centering
    \includegraphics[width=\linewidth]{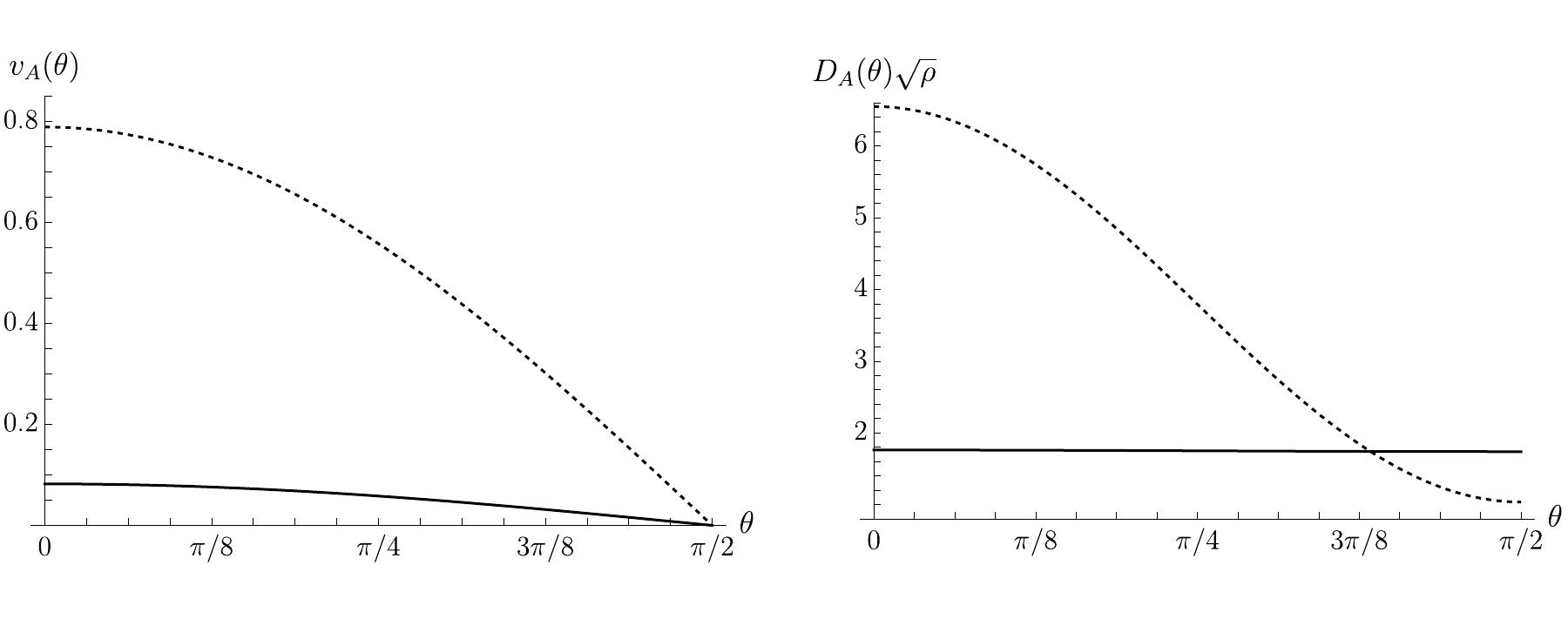}
    \caption{Speed of sound $v_A(\theta)$ (left) and diffusivity $D_A(\theta)$ (right) for Alfvén waves, calculated using the resistivities of the full theory. The solid line has $t\simeq 2$ (weak field regime), and the dashed one $t\simeq0.5$ (strong field regime).}
    \label{fig:alfven_tab}
\end{figure}

\paragraph{Transverse channel and full MHD}--- For simplicity, here, we only discuss the transverse channel of fluctuations in MHD. The hydrodynamic modes are Alfvén waves:
\begin{equation}
    \omega = \pm v_A \kappa - \frac{i}{2} \Gamma_A \kappa^2,
\end{equation}
where the speed of sound and attenuation are given by:
\begin{subequations}
    \label{eq:s5alf}
    \begin{align}
        & v_A = \sqrt{\frac{\mu \rho}{w}} \cos\theta, \\
        & \frac{1}{2} \Gamma_A = \pars{\frac{\eta_\perp}{w}+\frac{r_\perp}{\chi_\perp}} \sin^2\theta + \pars{\frac{\eta_\parallel}{w}+\frac{r_\parallel}{\chi_\perp}} \cos^2\theta.
    \end{align}
\end{subequations}
We see that no dependence on $\chi_\parallel$ enters the attenuation, in line with the findings of \cite{Grozdanov:2016tdf} (where $\chi_\parallel$ enters magnetosonic waves only). Additionally, the attenuation receives an imaginary part from the Drude peaks of the magnetic viscosities: for the sake of calculating `effective diffusivities', these may be ignored. To show the attenutation, we define $$D_A(\theta) = \frac{1}{2}\text{Re}\,\Gamma_A(\theta),$$ which we plot in \Cref{fig:alfven_tab}, along with the speed of sound $v_A(\theta)$. The approximately constant behavior of $D_A$ in the weak field regime can be understood by noticing that $r_\perp \approx r_\parallel$ and the suppression of the contribution from shear viscosities, which can be attributed to the large $w$ value. 

In both the transverse and longitudinal channels of MHD, the central difference between the probe limit and the full theory is in exhibiting diffusive versus sound-like behavior, respectively. In the physics of neutron stars, magnetic diffusion plays a central role \cite{goldreich1992magnetic,Vardhan:2022wxz}. It would therefore be of great phenomenological interest to understand under what physical conditions, the energy-momentum fluctuations become sufficiently important to induce a transition from diffusive to attenuated ballistic (Alfv\'{e}n and magnetosonic) propagation of magnetic and energy-momentum fluctuations. 

\section{Discussion}
\label{sec:disc}

At the beginning of this work, we set out to investigate the precise nature and validity the hydrodynamic probe limit in presence of an arbitrarily strong magnetic field, in a setting where electromagnetism is dynamical. In \Cref{sec:mhd}, under the lens of the EFT formulation of hydrodynamics as a derivative expansion, we showed how the probe limit, defined as the total decoupling of the evolution of the magnetic 2-form current $J^\mn$ from $T^\mn$, is valid in an approximate sense, violating the conservation of energy and momentum. This energy-momentum leakage is more clearly manifest in the failure of the canonical susceptibility matrix to block-diagonalize. Only a full theory of MHD can restore its conservation. We then proceeded with an analysis of these conclusions in a microscopic theory constructed via the holographic correspondence in \Cref{sec:grav_probe,sec:beyond}. We were able to substantiate the intuition that, while assuming the probe limit may lead to a theory that is perfectly defined, its predictions are often significantly different compared to the full theory. This is most clearly visible in presence of mixed correlators between $J^\mn$ and $T^\mn$ in the full theory (and the thermomagnetic effect), and the difference between diffusive and sound-like propagation. 

We see several possible future developments of this work. As discussed at the end of \Cref{sec:neutron}, it would be very interesting to better understand the regimes of neutron star physics, and other physical systems (see \cite{Vardhan:2022wxz,Vardhan:2024qdi}) in which the inclusion of full MHD becomes necessary. With respect to holography, our work advanced the understanding of the phenomenology captured by magnetic brane models, providing a dual perspective to earlier studies in the $n=0$ case (see \cite{Grozdanov:2017kyl,Grozdanov:2018fic,Li:2018ufq}), and providing the first calculation of a Hall electric transport coefficient in a gravitational theory with a 4-dimensional boundary (when $n\neq0$). With respect to this last point, one very simple extension of the calculation would be to include a scalar field to introduce momentum dissipation (as in \cite{Andrade:2013gsa,Baggioli:2021xuv}), or a massive vector field, making the model more realistic, with more intricate possibilities for discrete $C$, $P$ and $T$ symmetry patterns. Furthermore, one may try to substantiate further the intuition that the zero enthalpy point in the neutral theory corresponds to a phase transition. In particular, one may wonder how precisely this physics is related to the behavior of holographic superconductor. The analogy between our theory and ordinary charged hydrodynamics makes this point all the more interesting and worth investigating.

\acknowledgments{It is our pleasure to thank Napat Poovuttikul and Mile Vrbica for helpful discussions, and Arpit Das for collaboration in early stages of this project. The work of G.F. is supported by an Edinburgh Doctoral College Scholarship (EDCS). M.B. is supported by the research programme P1-0402 of Slovenian Research Agency (ARIS). G.K.B.~acknowledges support from the Slovenian Research Agency (ARIS) under the research programme N1-0392. The work of S.G. was supported by the STFC Ernest Rutherford Fellowship ST/T00388X/1. The work is also supported by the research programme P1-0402 and the project J7-60121 of Slovenian Research Agency (ARIS). A.S. was supported by funding from Horizon Europe research and innovation programme under the Marie Skłodowska-Curie grant agreement No. 101103006, the project N1-0245 of Slovenian Research Agency (ARIS) and financial support through the VIP project UNLOCK under contract no. SN-ZRD/22-27/510.}

\appendix

\section{Derivation of thermomagnetic transport matrix}\label{app:thermo-derivation}

In this appendix, we describe how to encode a temperature gradient in variations of the metric tensor, thus allowing us to obtain the thermomagnetic matrix in \eqref{eq:ThermoMagnetic_Matrix}. We mostly follow the treatment presented in Sec.~5.3 of \cite{Hartnoll:2016apf}.  

As appropriate for the scope of this work, we consider theories in flat space. It is convenient to work with a rescaled version of Euclidean time:
\begin{equation}
    ds^4_\ind{E} = \frac{1}{T_0^2} d\tau^2 + \delta_{ij} dx^idx^j, \quad \tau = - i T_0 t, 
\end{equation}
so that the the thermal circle has unit length $\tau \sim \tau + 1$. Now perturb the equilibrium temperature as in \eqref{eq:hvpt}, assuming that the perturbation has a constant spatial gradient and harmonic time dependence:
\begin{equation}
    T_0 \to T_0 \pars{ 1  + \delta \zeta_i x^i e^{-i \omegat \tau} },
\end{equation}
where we defined the dimensionless frequency $\omegat \equiv i \omega / T_0$ and gradient $\delta \zeta_i = \delta T_i / T_0$. This clearly induces a transformation on the metric tensor
\begin{equation}
    \eta_{\tau\tau} \to \eta_{\tau\tau} + \delta g_{\tau\tau}, \quad \delta g_{\tau\tau} \equiv - \frac{2}{T_0^2} \delta \zeta_i x^i e^{- \omegat \tau}.
\end{equation}
This transformation can be moved to a different metric component by acting via a suitable small diffeomorphism. These generically act on a covariant 2-tensor as:
\begin{equation}
    \label{eq:liegen}
    M_{\mu\nu} = M_\mn + (\mathcal L_\xi M)_\mn = \xi^\rho M_{\mu\nu,\rho} + {\xi^\rho}_{,\mu} M_{\rho\nu} + {\xi^\rho}_{,\nu} M_{\nu\rho},
\end{equation}
where $\xi$ is a vector field and $\mathcal L_\xi$ the Lie derivative along it. Using the symmetry of the metric tensor in \eqref{eq:liegen}, one is led to a simple first order ODE that determines the transformation needed to set $\delta g_{\tau\tau}=0$, which is given by
\begin{equation}
    \label{eq:xi_hls}
    \xi^\mu = \delta^\mu_\tau \frac{\delta\zeta_i x^i e^{-i \omegat \tau}}{- i \omegat}.
\end{equation}
This transformation acts on all components of the metric tensor, in particular making $\delta g_{\tau i}\neq 0$. Going back to real time, we find:
\begin{equation}
    \label{eq:TwithMetric}
    \delta g_{ti} = \frac{\partial \tau}{\partial t} \delta g_{\tau i} = \frac{\partial \tau}{\partial t} {\xi^{\tau}}_{,i} g_{\tau\tau} = \frac{1}{i\omega} \delta \zeta_i e^{-i\omega t} \quad\Rightarrow\quad \pars{\frac{\delta T}{T_0}}_i = \delta\zeta_{i} e^{-i\omega t} =  i \omega \delta g_{ti},
\end{equation}
which is used in the main body of this work. The transformation \eqref{eq:xi_hls} affects the 2-form gauge field source $b_{\mu\nu}$ as well. To analyze this case, it is most convenient to transform \eqref{eq:xi_hls} to real time:
\begin{equation}
    \label{eq:xi_hlst}
    \xi^\mu = \delta^\mu_t \frac{\delta\zeta_i x^i e^{-i \omega t}}{- i \omega}.
\end{equation}
We remind the reader that the equilibrium configuration is $b_\mn = \mu_0 u_{[\mu} h_{\nu]}$, so that the only nonzero background component is $b_{tz}$. The effect of the diffeomorphism \eqref{eq:xi_hlst} is that the fully spatial components of $b$ acquire a correction:
\begin{equation}
    \delta b_{ij} = {\xi^t}_{,i} b_{tj} - {\xi^t}_{,j} b_{ti} = (\delta \zeta_i b_{tj} - \delta\zeta_j b_{ti})\frac{e^{-i\omega t}}{-i\omega}.
\end{equation}
There is no spatial dependence in this expression, so that overall the correction to $H_{tij}$ reads:
\begin{equation}
    \delta H_{tij} = b_{ij,t} = (\delta \zeta_i b_{tj} - \delta\zeta_j b_{ti}){e^{-i\omega t}} = i\omega\mu_0 (\delta g_{ti}\delta^z_j- \delta g_{tj}\delta^z_i),
\end{equation}
as featured in the text.

\section{Hydrodynamic Ward identities}\label{app:WardMHD}

The hydrodynamic equations of motion
\begin{align*}
    \nabla_\mu J^\mn &= 0 \tag{\ref{eq:Jmn_Conservation}}, \\ 
    \nabla_\mu T^\mn &= H^{\nu}{}_\ab J^\ab,  \tag{\ref{eq:Tmn_Conservation}}
\end{align*}
imply the following Ward identities satisfied by the hydrodynamic correlators:
\begin{subequations}
\begin{align}
    &k_\mu G^{\mu\nu,\ab}_{JJ} = 0,\\
    &k_\mu G^{\mu\nu,\ab}_{JT} -k_\mu J^{\mu\nu}_\eq \eta^\ab=0,\\
    &k_\mu G_{TJ}^{\mu\nu,\ab}=-k^\nu J^\ab_\eq +k_\mu J^{\alpha\mu}_\eq \eta^{\beta\nu}-k_\mu J^{\beta\mu}_\eq \eta^{\alpha\nu},\\
    &k_\mu G^{\mu\nu,\ab}_{TT}-k_\mu T^{\mu\nu}_\eq \eta^\ab-k_\mu T^{\mu\beta}_\eq \eta^{\nu\alpha}-k_\mu T^{\mu\alpha}_\eq \eta^{\beta\nu}+k^\nu T^\ab_\eq=0,
\end{align}
\end{subequations}
with $T_\eq^\mn$ and $J^\mn_\eq$ denoting the background equilibrium values of energy momentum tensor and the conserved 2-form current, respectively. These identities yield extra constraints on the thermomagnetic transport matrix, defined in \eqref{eq:ThermoMagnetic_Matrix} and \eqref{eq:ThermoMag}. In the case of the background described in \Cref{sec:mhd}, given by \eqref{eq:ceqneq}, we find
\begin{subequations}
\label{eq:WardThermoMagnetic}
\begin{align}
    i\omega \left( T_\eq \tilde{\alpha}^{i,kl}+\mu_\eq h_\mu r^{i\mu,kl} \right) &= \rho_0 \left( h^k \eta^{li}-h^l \eta^{ki} \right),\label{eq:WardThermoMagnetic_1}\\
    i\omega \left( \bar{\kappa}^{i,k} + \mu_\eq h_\nu \alpha^{i\nu,k} \right) &= - s_0 \eta^{ik},\label{eq:WardThermoMagnetic_2}
\end{align}
\end{subequations}
where $T_\eq s_0 = \varepsilon_0 +p_0 - \mu_\eq \rho_0$. In the second equation we used the fact that the static susceptibility $\chi^{i,k}_{\bar{\kappa}} = \lim_{\omega\rightarrow 0}G_{QQ}^{i,k}$, is equal to $\chi^{i,k}_{\bar{\kappa}}  = -p_0\eta^{ik} +\mu_\eq \rho_0 h^i h^k$. Note that the index symmetry sets $\alpha^{zz,k} = 0$, implying
\begin{align}
    \bar{\kappa}^{z,z} = -\frac{s_0}{i\omega}.
\end{align}
Perhaps unsurprisingly, the constraints imposed on the thermomagnetic transport coefficients are highly reminiscent of the ones found in the case of ordinary (charged) hydrodynamics \cite{Hartnoll:2016apf,Hartnoll:2007ip,Herzog:2009xv}.

\section{Holographic renormalization and thermodynamic quantities}
\label{app:hrtq1}

Thermodynamic quantities such as the energy density $\varepsilon$ and pressure $p$ can be holographically extracted by an asymptotic analysis of the metric. This is usually done by adopting Fefferman-Graham (FG) coordinates. For neutral magnetic branes, this analysis has been performed in \cite{Grozdanov:2017kyl}. Because of the asymptotic order at which $n$ enters the metric coefficients, the expressions are formally the same as in the $n\neq0$ case. The purpose of the present discussion is to clarify the relationship between FG coordinates and the physical $q$ ones.  

Our choice of FG coordinates is the one adopted in \cite{Skenderis:2002wp}:
\begin{align}
    & ds^2 = \frac{d\rho^2}{4\rho^2} + \frac{1}{\rho}g_{\mu\nu}(\rho,x)dx^\mu dx^\nu, \nonumber \\[4pt]
    \label{ae:fgexp}
    & g_{\mu\nu}(\rho,x)\xrightarrow{\rho\to0} g^{(0)}_\mn(x) + \rho \, g^{(1)}_\mn(x) + \rho^2 g^{(2)}_\mn(x) + \rho^2 \ln\rho \, \tilde h_\mn(x) + ...
\end{align}
The presence of a logarithmic term in the expansion is typical of gravitational bulk theories with an odd number of spacetime dimensions. In particular, $\tilde h_{\mu\nu}\neq0$ signals the breakdown of conformal symmetry on the boundary, which in our case can be precisely traced to the presence of a background magnetic field (as pointed out in earlier works such as \cite{Fuini:2015hba}).

A peculiarity of the 2-form magnetic brane model \eqref{eq:uberS} is that the coefficient multiplying the logarithmic term in the matter field $B_{ab}$ is free:
\begin{equation}
    B_{\mu\nu}(\rho,x) \xrightarrow{\rho\to0} B_\mn^{(0)}(x) + \ln \rho \, B_\mn^{(1)}(x) + ...
\end{equation}
Because of this, the on-shell action has a logarithmic divergence, the regularization of which is inherently ambiguous. In \cite{Grozdanov:2017kyl}, this ambiguity was interpreted as the choice of renormalization scale of the theory. In this picture, the prescription to remove all divergences from the on-shell action appears to be simply one of infinitely many choices. This is reflected in the expectation value of $T_{\mu\nu}$, which reads:
\begin{equation}
    \label{eq:aemt}
    \thev{T_{\mu\nu}} = g^{(2)}_\mn - g_\mn^{(0)} {(g^{(2)})^\alpha}_\alpha + \frac{1}{2} \tilde h_\mn - \frac{1}{2\pi\bar\alpha} \tilde h_\mn.
\end{equation}
The usual holographic renormalization procedure corresponds to $\bar\alpha\to\infty$, whereas in \cite{Grozdanov:2017kyl} $\bar\alpha=1/137$. We refer the interested reader to that work for an extensive discussion of the topic.

Importantly, the coefficients in \eqref{eq:aemt} are the ones appearing in the FG expansion \eqref{ae:fgexp}, whereas we can access coefficients of the asymptotic expansions in $q$ (and know how to translate between physical and numerical coordinates within that framework). To find the precise relationship between FG and numerically accessible coefficients, we need to solve the following differential equation:
\begin{equation}
    \frac{d\rho}{2\rho} = \frac{dq}{q\sqrt{u(q)}},
\end{equation}
which close to the boundary leads to:
\begin{align}
    & \rho \simeq q^2 + \pars{-\frac{u_{(4)}}{4} + \frac{\rho_\text{phys}^2}{24}} q^6 - \frac{1}{6}\rho_\text{phys}^2 q^6 \ln q - \frac{n_\text{phys}^2}{72} q^8 + ..., \nonumber \\
    & \frac{1}{\rho} \simeq \frac{1}{q^2} + \pars{\frac{u_{(4)}}{4} - \frac{\rho_\text{phys}^2}{24}} q^2 + \frac{1}{6}\rho_\text{phys}^2 q^2 \ln q + \frac{n_\text{phys}^2}{72} q^4 ... .
\end{align}
These relationships can be used to express the energy-momentum tensor \eqref{eq:aemt} with $q$-asymptotic UV coefficients:
\begin{equation}
    \label{eq:aemtq}
    \thev{T_{\mu\nu}} = g^{(4)}_\mn - g_\mn^{(0)} \pars{{(g^{(4)})^\alpha}_\alpha - \frac{3}{4} u_{(4)}+ \frac{\rho^2_\text{phys}}{6} -\frac{\rho^2_\text{phys}}{24 \pi \bar\alpha}} +\frac{1}{4} \tilde h_\mn \pars{1 - \frac{1}{\pi\bar\alpha}}.
\end{equation}
Once the rescaling to physical coordinates \eqref{eq:cntp} is taken into account, we can read from \eqref{eq:UVb_expFG} the following relations:
\begin{equation}
    \begin{aligned}
        g^{(0)}= \eta,\quad g^{(4)}=\text{diag}(-u_{(4)},2v_{(4)},2v_{(4)},-4v_{(4)}), \quad \tilde h=\frac{\rho^2_\text{phys}}{3}\, \text{diag}(-2,1,1,2).
    \end{aligned}
\end{equation}
Then substitution into \eqref{eq:aemtq} leads to:
\begin{subequations}
    \begin{align}
        & \varepsilon = \thev{T_{tt}} = -\frac{3}{4}u_{(4)}+\frac{\rho^2_\text{phys}}{8\pi\bar\alpha}, \\
        & p = \thev{T_{aa}}\Big|_{a=x,y} = 2v_{(4)} -\frac{1}{4}u_{(4)}+\frac{\rho^2_\text{phys}}{4} \pars{\frac{1}{2\pi\bar\alpha}-1}, \\
        & p - \mu\rho = \thev{T_{zz}} = -4 v_{(4)} -\frac{1}{4}u_{(4)} - \frac{\rho^2_\text{phys}}{8\pi\bar\alpha},
    \end{align}
\end{subequations}
which are reported in the main text. While superficially different, these formulae are in complete agreement with the ones given in \cite{Grozdanov:2017kyl} once the different choice of coordinates and UV expansion of the metric components is taken into account. As we remarked at the beginning, the expression \eqref{eq:aemtq} is only superficially independent of $n_\text{phys}$, since all boundary coefficients -- which are found via numerical integration of the ODEs in the bulk -- depend on it.

\bibliography{main}

\bibliographystyle{jhep}

\end{document}